\documentclass[pra,superscriptaddress,twocolumn]{revtex4-1} 
\usepackage{graphicx,amsfonts,bm,amsmath,mathrsfs,amssymb}
\usepackage[usenames,dvipsnames]{color}

\def \v#1{{\bm #1}}
\def \be {\begin{equation}}
\def \ee {\end{equation}}
\def \bml{\begin{multline}}
\def \eml{\end{multline}}

\newcommand{\I}{\mathrm{i}}

\newcommand{\tr}[1]{\mathrm{Tr}\{#1\}}
\newcommand{\Tr}[1]{\mathrm{Tr}\Big\{#1\Big\}}
\newcommand{\Det}[1]{\mathrm{Det}\{#1\}}

\def \del{\partial}

\def \ds{\displaystyle}

\def \DEF{\stackrel{\rm def}{\Leftrightarrow}}
\def \Lra{\Leftrightarrow}

\def \sld{\mathrm{SLD}}

\def \nnd{\mathrm{NND}(n)}
\def \pd{\mathrm{PD}(n)}

\def \cE{{\cal E}}
\def \cJ{{\cal J}}
\def \cM{{\cal M}}
\def \cH{{\cal H}}
\def \cP{{\cal P}}
\def \cT{{\cal T}}

\def \cX{{\cal X}}

\def \PsetX{{\cal P}(\cX)}
\def \sofh{{\cal S}({\cal H})}

\def \bbc{{\mathbb C}}
\def \bbn{{\mathbb N}}
\def \bbr{{\mathbb R}}

\renewcommand{\vec}[1]{\mathbf{#1}}

\newcommand{\param}[1]{\theta_{#1}}
\newcommand{\hparam}[1]{\hat{\theta}_{#1}}
\newcommand{\Param}{\Theta}
\newcommand{\vparam}[1]{\vartheta_{#1}}
\newcommand{\hvparam}[1]{\hat{\vartheta}_{#1}}

\newcommand{\qfi}[1]{J_{#1}^{SLD}}
\newcommand{\odoe}{optimal DoE}

\newcommand{\qed}{\nobreak \ifvmode \relax \else
	\ifdim\lastskip<1.5em \hskip-\lastskip
	\hskip1.5em plus0em minus0.5em \fi \nobreak
	\vrule height0.75em width0.5em depth0.25em\fi}

\begin{document}
\title{Quantum process tomography via optimal design of experiments}
	
\author{Yonatan Gazit}
\affiliation{Yale-NUS College, Singapore}
\author{Hui Khoon Ng}
\affiliation{Yale-NUS College, Singapore}
\affiliation{Centre for Quantum Technologies, National University of Singapore}
\affiliation{MajuLab, International Joint Research Unit UMI 3654, CNRS, Universit{\'e} C{\^o}te d'Azur,
Sorbonne Universit{\'e}, National University of Singapore, Nanyang Technological University, 
Singapore}
\author{Jun Suzuki}
\affiliation{Graduate School of Informatics and Engineering, The University of Electro-Communications, Tokyo, Japan}
\date{\today}
	
\begin{abstract}
Quantum process tomography---a primitive in many quantum information processing tasks---can be cast within the framework of the theory of design of experiments (DoE), a branch of classical statistics that deals with the relationship between inputs and outputs of an experimental setup. Such a link potentially gives access to the many ideas of the rich subject of classical DoE for use in quantum problems.
The classical techniques from DoE cannot, however, be directly applied to the quantum process tomography due to the basic structural differences between the classical and quantum estimation problems. Here, we properly formulate quantum process tomography as a DoE problem, and examine several examples to illustrate the link and the methods. In particular, we discuss the common issue of nuisance parameters, and point out interesting features in the quantum problem absent in the usual classical setting.
\end{abstract}
	
\keywords{parameter estimation, 
quantum process tomography, optimal design of experiment, Cram\'er-Rao type bound
}
	
\maketitle
	
\section{Introduction}\label{sec1}

Design of experiments (DoE) is a branch of mathematical statistics that examines efficient methods to understand the relationship between inputs and outputs for an experimental setup. Founded by R. A. Fisher in the early $20^{\text{th}}$ century, further developments in DoE were made by several mathematical statisticians like Wald, Kiefer, Chernoff, and Fedorov, to name a few. One of the celebrated results is the so-called equivalence theorem put forth by Kiefer and Wolfowitz \cite{kw60}, which established equivalence among different optimal designs. 
	
The problem of quantum process (or channel) tomography---a primitive in many quantum information processing tasks---can be considered as an optimal design problem. Here, the goal is to estimate the quantum process/channel, considered here as a black box that takes in an input quantum state and puts out a modified output state. The experimenter chooses the input probe states and decides on how to measure the outputs to obtain a description of the inner workings of the black box. One can naturally formulate the problem of estimating a parametric family of quantum channels based on the theory of \odoe{}, permitting the application of the established machinery of classical statistics to finding optimal quantum tomography strategies. 
	
It turns out, however, that many of the classical techniques from DoE cannot be directly applied to the quantum problem in a straightforward manner. A major obstacle is the differences in the structure of the state and measurement spaces of quantum estimation problems compared to the classical case. Moreover, many of the previous \odoe{} studies in statistics were carried out for the linear regression model and its variants, inapplicable to tomography problems in quantum systems. Here, we extend methodologies for non-linear models studied in Refs.~\cite{fedorov,pukelsheim,fh97,fl14,pp13} 
to the more general formulation of \odoe{}, which is applicable for any probabilistic model. 
	
There are already previous attempts to apply the theory of \odoe{} to quantum tomography. This was first done in Ref.~\cite{kwr04}, and following that study, several more papers on the subject appeared over the past decade \cite{nunn10,bh10,bh11,bhp12,rvh12}. These studies, however, dealt only with limited cases, e.g., discrete design problems, or the optimization of the relative frequencies for different experimental settings. The former setting is practically important, yet it is hard in general hard to get the solution even numerically; the latter setting was analyzed under the assumption of given tomographic measurement settings. 
	
Here, we formulate the problem of quantum process tomography in the most general setting, including the discrete design problem as a special case, and address also the common issue of nuisance parameters. In Sec.~\ref{sec2}, we provide a review of the relevant concepts in the classical theory of \odoe{}, necessary to familiarize the reader with the basic ideas. We reformulate those ideas in the quantum setting in Sec.~\ref{sec3}. In Sec.~\ref{sec4}, we elucidate different features of the problem through examples for qubit models, and illustrate the usefulness of the \odoe{} framework in quantum tomography problems. 
Supplemental materials about the classical theory of \odoe{} are given in the Appendix.  

Note that we focus only on tomography strategies that do not require expensive resources such as entangled states, or the ability to perform join measurements \cite{fujiwara01,fi03,glm11}. Of course, the use of such additional resources gives higher performance in general, but practically, entangled resources and joint operations remain difficult to achieve in the lab today. The \odoe{} approach requires only control over the input probe states and output measurements. 
The formulation presented in this paper can easily be extended to more general settings that make use of these quantum mechanical resources.

\section{Classical theory of \odoe{}}\label{sec2}
In this section, we provide a brief summary of the classical theory of \odoe{} developed along the lines of Refs.~\cite{fedorov,pukelsheim,fh97,fl14,pp13}. To simplify matters, we focus on point estimation problems about parameter models and probability distributions on discrete sets. Other statistical inference problems, such as hypothesis testing, model discrimination, and so on, can be formulated in a similar manner.

\subsection{Local optimal design}\label{sec2-1}
An $n$-parameter coordinate system is denoted by ${\param{}}=(\param{1},\param{2},\dots,\param{n})$ to describe the object of interest. The parameter ${\param{}}=(\param{i})$, the {\it model parameter}, takes values in $\Param$, an open subset of $\bbr^n$. Let us introduce a {\it design} $e$ describing a particular experimental setup, and let $\cE$ be the set of all possible designs. 
A {\it model function} $f$ is a mapping from $\Param\times\cE$ to a set of probability distributions on $\cX (\equiv \PsetX$), that is, $f:\, (\param{},e)\mapsto p_{\param{}}(\cdot|e)\in\PsetX$ where, $\forall x\in\cX$, $ p_{\param{}}(x|e)\ge0$ and $\sum_{x\in\cX}p_{\param{}}(x|e)=1$. 
Note that the concept of a model function is not introduced in the classical \odoe{}. 
But this is essential for extending the formalism of linear regression models to the more general probabilistic models.
	
We assume that the model set $\Param$ is continuous. The design set $\cE$, on the other hand, can be arbitrary, and is determined by the given experimental configuration or constraints. The element $e\in\cE$ can be a vector, a matrix, or a more general object (see concrete examples below).
	
Given an unknown object smoothly parametrized by $\param{}$, we choose a proper design $e$ that gives a particular statistical model,
\begin{equation} \label{eq:stat.model}
M(e)= \{p_{\param{}}(\cdot|e)\,|\, \param{}\in\Param\},
\end{equation} 
according to a known model function $f$. The experimental data $X$ is a random variable defined by $p_{\param{}}(\cdot|e)$. The value of $ \param{} $ is inferred from some data $x\in\cX$ by using an estimator ${\hparam{}:\cX\to\Param}$, ${\hparam{}=(\hparam{1},\dots,\hparam{n})}$. We use the mean-square error (MSE) matrix, a non-negative $n\times n$ real matrix, as a measure of an estimator's error. Let ${E_{\param{}}[X|e]= \sum_{x\in\cX} x p_{\param{}}(x|e) }$ be the expectation value of a random variable $X$ with respect to $p_{\param{}}(\cdot|e)$. 
The MSE matrix is defined by
\begin{equation}\label{eq:mse.matrix}
V_{\param{}}[\hparam{}|e]=\Big[ E_{\param{}}[(\hparam{i}-\param{i}) (\hparam{j}-\param{j})|e] \Big]_{i,j}.
\end{equation}
When reconstructing the value of $ \param{} $ from the data, the goal is to find the estimator that minimizes the MSE matrix for a design $e\in\cE$. 
	
As is well known, there cannot in general be a universally optimal estimator that minimizes the MSE matrix for all $\param{}\in\Param$ \cite{rao73,kiefer87,lc98}. We thus look for an optimal estimator within a subclass of estimators. In this paper, we consider only locally unbiased estimators, defined as follows: An estimator $\hparam{}$ is said to be {\it locally unbiased at $\param{0}$} for a design $e\in\cE$ if $E_{\param{0}}[\hparam{i}|e]=\param{i}$ and $\frac{\del}{\del\param{j}}E_{\param{0}}[\hparam{i}|e]\big|_{\param{} = \param{0}}=\delta_{i,j}$ are satisfied for $\forall i,j$ at a particular point $\param{0}$. 
	
We can now make use of the Cram\'er-Rao theorem \cite{rao73,kiefer87,lc98} for a  fixed design $e$, assuming that the model $M(e)$ satisfies the usual regularity conditions. 
The well-known Cram\'er-Rao (CR) theorem states that the MSE matrices for all locally unbiased estimators are bounded by
\begin{equation} \label{eq:crb}
V_{\param{}}[\hparam{}|e]\ge \Big( J_{\param{}}[e]\Big)^{-1}. 
\end{equation}
Here $J_{\param{}}[e]$ is the Fisher information matrix about the statistical model $M(e)$ for the design $e$, defined as
\begin{equation} \label{eq:cl.fish}
J_{\param{}}[e]= \Big[ E_{\param{}}[ \frac{\del \ell_{\param{}}(X|e)}{\del\param{i}}  \frac{\del\ell_{\param{}}(X|e)}{\del\param{j}}  \Big| e] \Big]_{i,j}, 
\end{equation}
where $\ell_{\param{}}(x|e)=\log p_{\param{}}(x|e) $ is the logarithmic likelihood function. 
Importantly, the above CR inequality can be saturated asymptotically (i.e., in the sample size $N\to\infty$ limit). 
An optimal design $e$, therefore, maximizes the Fisher information matrix $J_{\param{}}[e]$.
	
However, it is usually impossible to minimize a matrix function as a matrix inequality because the matrix ordering does not yield a totally ordered set for all information matrices. In such cases, one has to adopt some other suitably chosen optimal criteria. These optimal criteria can be expressed in terms of an {\it optimality function} $ \Psi $, a function of non-negative matrices such that $\Psi(A)\ge0 $ for all $A\ge0$. We can then formulate the optimization problem in terms of the chosen optimality function $ \Psi $:
\begin{align*}
\Psi_*&=\min_{e\in\cE} \Psi\Big(J_{\param{}}[e]\Big),\\ 
e_*&=\mathrm{arg}\min_{e\in\cE} \Psi\Big(J_{\param{}}[e]\Big). 
\end{align*}
The optimal design $e_*$ is said to be {\it $\Psi$-optimal}. 
	
In the theory of \odoe{}, there are various optimality criteria commonly used to define the best design. We list below some standard criteria by which to define an optimality function $ \Psi $ (see Appendix Sec.~\ref{sec-app_supp1} for supplemental material and Refs~\cite{pukelsheim,fh97,fl14,pp13} for more details). 
	\begin{itemize}
		\item \underline{L\"owner optimality $ \Psi_{L} $:} \\
		$e_*$ is L\"owner optimal 
		$\DEF$ $\exists e_*\in\cE$ such that\\
		\hfill $\forall e\in\cE\, J_{\param{}}[e]\le J_{\param{}}[e_*]$ 
		and $\exists e',\,J_{\param{}}[e']< J_{\param{}}[e_*]$.
		
		\item \underline{$A$-optimality $ \Psi_{A} $:} \\
		$e_*$ is $A$-optimal $\DEF$ $e_*=\arg\min \Tr{J_{\param{}}[e]^{-1}}$.
		
		\item \underline{$D$-optimality $ \Psi_{D} $:} \\
		$e_*$ is $D$-optimal $\DEF$ $e_*=\arg\min \Det{J_{\param{}}[e]^{-1}} $ \\
		\hspace{2.5cm}$\Lra$ $e_*=\arg\max \Det{J_{\param{}}[e]} $.
		
		\item \underline{$E$-optimality $ \Psi_{E} $:} \\
		$e_*$ is $E$-optimal 
		$\DEF$ $e_*=\arg\min \lambda_{\max}({J_{\param{}}[e]^{-1}}) $\\
		\hspace{2.47cm}$\Lra$ $e_*=\arg\max \lambda_{\min}({J_{\param{}}[e]}) $ ,\\
		where $\lambda_{\max}(A)$ and $\lambda_{\min}(A)$ are the largest and smallest eigenvalues, respectively, of a symmetric matrix $A$. 
		
		\item \underline{$c$-optimality $ \Psi_{c} $: }\\
		$e_*$ is $c$-optimal $\DEF$ $\ e_*=\arg\min c^\mathrm{T}J_{\param{}}[e]^{-1} c$,\\
		where $c\in\bbr^n$ is a given column vector. 
		
		\item \underline{$\gamma$-optimality $ \Psi_{\gamma} $:} \\
		$e_*$ is $\gamma$-optimal ($\gamma\in(0,\infty)$)\\
		\qquad$\DEF$ $e_*=\arg\min \left(\frac1n\Tr{J_{\param{}}[e]^{-\gamma}}\right)^{1/\gamma}$,\\
		where $n$ is the dimension of the parameter set $\Param$. 
	\end{itemize}
	The \mbox{$A$-,} $D$-, and $E$-optimal designs are named as the average optimal design, the optimal about the determinant, and the optimal about the extremal eigenvalue, respectively. 
	
We list some terminology concerning designs below.
	If an optimal design $e_*$ is a function of the unknown parameter(s) $\param{}$, it is called a {\it local optimal design} in the sense that it is optimal at a specific point $\param{0}$. 
	Without {\it a priori} knowledge about $\param{}$, it is impossible to immediately perform this optimal design $e_*$, but there exist various sequential algorithms realizing $e_*$ in the sample size $N\to\infty$ limit. 
	On the other hand, when $e_*$ is $\param{}$-independent, it is called a {\it globally optimal design}. 
	A well-known example of a globally optimal design is for the linear regression model, where the optimal design is always $\param{}$ independent.
	Alternatively, one can look for an averaged optimal design, a Bayesian optimal design, or a min-max optimal design to avoid $\param{}$ dependence in $e_*$, see Refs.~\cite{fedorov,pukelsheim,fh97,fl14,pp13}. 
	In this paper, we mainly focus on local optimal designs.
Another terminology concerns singular behavior of designs. When the Fisher information matrix $J_{\param{}}[e]$ 
is not full rank for a design $e$, we say $e$ is a {\it singular design}. The singular design problem is discussed in Appendix Sec.~\ref{sec-app_sing}. 
	
A few remarks about the optimality criteria are in order. 
First, $A$-optimality can be generalized to minimizing $\Tr{WJ_{\param{}}[e]^{-1}}$, where $W\ge0$ is a non-negative matrix, called a weight matrix, utility matrix, or loss matrix. 
	The introduction of an appropriate $ W $ allows one to focus on the parameters of interest, and this formulation is often adopted in parameter estimation of quantum states. 
	The L\"owner optimal design is the strongest criterion in the sense that if there exists a L\"owner optimal design $e_*$, then all other optimality criteria are automatically satisfied. However, this occurs only for very special models.
	We elaborate on this point in Appendix Sec.~\ref{sec-app_lowner}.
	The $\gamma$-optimality criterion contains the $A$-optimal ($\gamma=1$), $D$-optimal ($\gamma\to0$), and $E$-optimal ($\gamma\to\infty$) criteria as special cases. But a closed expression for the $\gamma$-optimal design is hard in general to obtain. (See also Appendix Sec.~\ref{sec-app_supp1}.)

\subsection{Discrete and continuous design problems}\label{sec2-5}

	In this subsection, we extend our discussion to multiple design problems. 
	When considering a situation of $N$ repetitions of an experiment, there are two distinct strategies to choose from:\\[1ex]
\noindent i) \underline{i.~i.~d.~strategy}. Repeat the same design $e$ for a total of $N$ times. 
	Let us refer to the design of this strategy as $e^N\in\cE^N$. 
	The probability distribution for model becomes an independently and identically distributed (i.~i.~d.) one,  
	\begin{equation*}
	p_{\param{}}(x^N|e^N)=\prod _{t=1}^N p_{\param{}}(x_t|e),
	\end{equation*}
	because of the additivity of the Fisher information matrix, $J_{\param{}}[e^N]=NJ_{\param{}}[e]$. 
	The problem is solved by considering the $N=1$ case. \\[1ex]
	\noindent
	ii) \underline{Mixed strategy}. Let $N(m)$ be an $m$-partition of a positive integer $N$, i.e., $N(m)=(n_1,n_2,\dots,n_m)$ such that $\sum_{i=1}^mn_i=N$ and $n_i\ge0$. 
	The mixed strategy involves repeating a design $e_1$ for $n_1$ times, $e_2$ for $n_2$ times, and so on, for all $m$ designs. 
	Let us refer to this strategy's design as $e[N(m)]$. The probability distribution is then
	\begin{equation*}
	p_{\param{}}\bigl(x^N|e[N(m)]\bigr)=\prod _{i=1}^{m} p_{\param{}}(x^{n_i}|e_{i}^{n_i})=\prod _{i=1}^{m} \prod_{t_i=1}^{n_i}p_{\param{}}(x_{t_i}|e_{i}), 
	\end{equation*}
	and the Fisher information matrix for $e[N(m)]$ is 
	\begin{equation}
	J_{\param{}}\big[e[N(m)]\big]=\sum_{i=1}^m n_i J_{\param{}}[e_i]. 
	\end{equation}
	When $N$ is fixed, the optimization corresponds to finding the partition $N(m)$ that minimizes the optimality function $\Psi\bigl(J_{\param{}}\big[e[N(m)]\big]^{-1}\bigr)$. 
	This optimization is known as a {\it discrete design} or {\it exact design} problem. 
	In the very special situation where a L\"owner optimal solution exists, an optimal mixed strategy corresponds to the i.~i.~d.~strategy. 
	
Although the combinatoric optimization of a discrete design problem is practically important, it is in general hard to find an optimal solution, even numerically. The standard approach to finding an approximate optimal solution is to consider instead a {\it continuous design} problem (also known as an {\it approximate design} problem).
Taking the $N\to\infty$ limit, the normalized proportions become relative frequencies, $\nu_i=\lim_{N\to\infty} (n_i/N)$. The goal is then to find the optimal relative frequencies $\v{\nu}=(\nu_i)\in\cP(m)$ and the set of designs $\v{e}=(e_i)\in\cE^m$ that minimize $\Psi\bigl(J_{\param{}}\big[e(m)]\big]^{-1}\bigr)$. Here, we denote the design of this continuous design problem by 
\begin{align*}
e(m)&=(\v{\nu},\v{e})\in\cP(m)\times\cE^m\\
&=\Big((\nu_1,\dots,\nu_m),\,(e_1,\dots,e_m) \Big).
\end{align*}
The Fisher information matrix for the design $e(m)$ is then
\begin{equation*}
J_{\param{}}[e(m)]=\sum_{i=1}^m \nu_iJ_{\param{}}[e_i]. 
\end{equation*}
This is equivalent to the Fisher information of the joint probability distribution $\sum_i \nu_ip_{\param{}}(x|e_i)$.  
To phrase it differently, the mixed strategy amounts to maximizing the Fisher information for the statistical model
\begin{equation}
M\left(e(m)\right)={\left\{\sum_{i=1}^m \nu_ip_{\param{}}(\cdot|e_i)\,|\,\param{}\in\Param\right\}}.
\end{equation}
	
The continuous design problem can be summarized as follows: Given an optimality function $ \Psi $ and a positive integer $m$, one must find an optimal design $e_*(m)=(\v{\nu}_*,\v{e}_*)$ defined by 
\begin{equation}\label{eq:opt.design}
e_*(m)=\arg\hspace{-5mm}\min_{e(m)\in\cP(m)\times\cE^m} \Psi\bigg(\sum_{i=1}^m \nu_iJ_{\param{}}[e_i]\bigg). 
\end{equation}
We plan to find an optimal value for $ m $ by sequentially finding the optimal design for different values of $ m $. 
That is, for some fixed $ m $ we find $ e_{*}(m),e_{*}\mbox{(m+1)},e_{*}(m+2), $ and so on. 
By comparing the optimal designs of various $m$ values, we can search for the optimal $e_*(m_*)$ over all possible designs. The general theorem (Carath\'eodory's theorem) guarantees that an optimal design can be found by using no more than \mbox{$n(n+1)/2+1$} designs, where $n$ is the number of parameters to be estimated \cite{fedorov,pukelsheim,fh97,fl14,pp13}. In the presence of $\ell$ independent constraints on the design $e$, this upper bound becomes $n(n+1)/2+\ell$ \cite{fh97,fl14}.
	
Before closing this section, we have one remark. From the expression in Eq.~\eqref{eq:opt.design}, it is clear that a closed expression for the optimal continuous design cannot be obtained except in special cases. Therefore, we often have to use numerical search instead to find the optimal design. This has also been an area of active research in the field of \odoe{} \cite{pukelsheim,fh97,fl14,pp13}.

\subsection{Nuisance Parameters}\label{sec2-3}
For an $n$-parameter object, often only $k<n$ parameters are of interest. The parameters not of interest are called {\it nuisance parameters} in statistics. The nuisance parameter problem is very important in many areas of statistics and have been studied since Fisher's work in 1935 \cite{fisher35}. In classical statistics, there are various methods to eliminate nuisance parameters and find a good estimator for the parameters of interest; see, for example, textbooks \cite{amari85,lc98,bnc94,an00} and Refs.~\cite{basu77,rc87,ak88,bs94,zr94}.
	
We can formulate the nuisance parameter problem by dividing an $n$-parameter object into two groups $\param{}=(\param{I},\param{N})$.
$\param{I}=(\param{1},\param{2},\dots,\param{k})$ are the parameters of interest and $\param{N}=(\param{k+1},\param{k+2},\dots,\param{n})$  are the nuisance parameters. 
Our aim is then to find a good design $e$ and to construct a good estimator $\hparam{I}=(\hparam{1},\hparam{2},\dots,\hparam{k})$ for the parameters of interest. 
	
Let us decompose $J_{\param{}}$ and $J_{\param{}}^{-1}$ into block matrices according to the parameter group $\param{}=(\param{I},\param{N})$:
\begin{align}
J_{\param{}}&=\left(\begin{array}{cc}J_{\param{},II} & J_{\param{},IN} \\ J_{\param{},NI} & J_{\param{},NN}\end{array}\right),\\
J_{\param{}}^{-1}&=\left(\begin{array}{cc}J_{\param{}}^{II} & J_{\param{}}^{IN} \\[0.5ex] J_{\param{}}^{NI} & J_{\param{}}^{NN}\end{array}\right),
\end{align}
where we have dropped the  $ e $-dependence in the notation. 
$(J_{\param{}}^{II}[e])^{-1}$ is called the {\it partial Fisher information} about the model \cite{zr94}, 
as its inverse $J_{\param{}}^{II}[e]$ provides a bound for the estimation error about the parameters of interest.
The MSE matrix $V_{\param{I}}$ for the parameters of interest is a $k\times k$ real symmetric matrix and is bounded by the partial Fisher information, 
\begin{equation}\label{eq:CRnuisance}
V_{\param{I}}[\hparam{I}|e]\ge J_{\param{}}^{II}[e]. 
\end{equation}
Using standard matrix analysis, the partial Fisher information can be expressed as
\begin{equation}
(J_{\param{}}^{II}[e])^{-1} = J_{\param{},II} - J_{\param{},IN}J_{\param{},NN}^{-1}J_{\param{},NI}.
\end{equation}
Therefore, $(J_{\param{}}^{II}[e])^{-1}\le J_{\param{},II}$, with equality if and only if  $J_{\param{},IN}=0$. 
If all nuisance parameters $\param{N}$ are known, then the problem is reduced to a $k$-parameter estimation problem and the CR inequality for $\param{I}$ becomes 
\begin{equation}\label{eq:CRnuisance2}
V_{\param{I}}[\hparam{I}|e]\ge (J_{\param{},II}[e])^{-1}. 
\end{equation} 
Comparing the CR inequality in Eqs.~(\ref{eq:CRnuisance}) and (\ref{eq:CRnuisance2}), we can conclude that the two lower bounds are the same if and only if $J_{\param{},IN}=0$. 
Otherwise, we cannot ignore the effect of nuisance parameters in estimating the parameters of interest. 
The presence of nuisance parameters leads to a larger lower bound for the error.  
	
Here, we will only concern ourselves with the $A$-optimal design when it comes to nuisance parameter estimation problems. 
We modify the optimality function to be $\Psi_{W}(J)=\Tr{WJ^{-1}}$ and set 
	the weight matrix as 
	\begin{equation}
	W=\left(\begin{array}{cc}W_{I} & 0 \\ 0& 0\end{array}\right). 
	\end{equation}
	where $W_{I}$ is a $k\times k$ positive matrix. 	When $W_{I}=I_k$ (the identity matrix for the $k\times k$ sub-block), $\Psi_{W}(J)  $ is equivalent to optimizing $\Tr{J_{\param{}}^{II}[e]}$, the $A$-optimality function for the parameters of interest. 
	Similar extensions can be done to define other optimal designs in the presence of nuisance parameters.

\section{Quantum channel parameter estimation}\label{sec3}
Building upon the last section, we now connect the theory of \odoe{} to quantum process tomography. More specifically, we discuss \odoe{} for estimating the parameters of a given family of quantum processes, also known as quantum channels. We first list several definitions (axioms) of quantum system (see, for example, Ref.~\cite{NC,petz} for details) before formulating \odoe{} in a quantum setting.

\subsection{Definitions}
\noindent
Q1) \underline{System}. A {\it quantum system} is represented by a $d$-dimensional complex vector space $\bbc^d$. 
	With the standard inner product, it becomes a Hilbert space denoted by $\cH=\bbc^d$. 
	When the dimension of the system is two, we speak of ``qubit", the simplest quantum system. 
	To simplify our discussion we only consider quantum systems with a fixed dimension $d<\infty$.\\[1ex]
\noindent
Q2) \underline{States}. A {\it quantum state} is represented by a non-negative matrix $\rho$ on $\cH$ with unit trace. 
	The set of all quantum states on $\cH$ is denoted by $\sofh=\{\rho\,|\,\rho\ge0,\tr{\rho}=1 \}$.
	When we analyze the qubit problem, a convenient representation of qubit states is as follows. 
	Define a bijective map from a $2\times2$ Hermitian matrix $A\in\bbc^{2\times2}$ 
	($A^\dagger=A$, where $A^\dagger$ denotes the Hermitian conjugate of a complex matrix $A$)
	to a three-dimensional real vector $\vec{s}=(s_i)$ via $s_i=\tr{\rho\sigma_i}$ where $\sigma_i$ ($i=1,2,3$) 
	are the Pauli matrices \cite{pauli}. $\rho$ is a physical quantum state if and only if $|\vec{s}|\le1$. This real vector is referred to as the {\it Bloch vector representation} of the state $\rho$. 
	A state with a Bloch vector of unit length, i.e., $|\vec{s}|=1$, is referred to as a {\it pure state}, corresponding to the situation where the quantum system is in a definite state. Pure states are the extremal points of the convex state space $\sofh$. \\[1ex]
\noindent
Q3) \underline{Measurements}. 
	A {\it measurement} $\Pi$ on a given quantum state $\rho$ is described as a set of non-negative matrices 
$\Pi=\{\Pi_x\}_{x\in\cX}$ such that $\sum_{x\in\cX}\Pi_x=I_d$, where $\cX$ is the index set of all the measurement outcomes $\Pi_x$s. 
	The probability of observing the measurement outcome represented by $\Pi$ for a state $\rho$ is given by the {\it Born's rule}, which defines the model function,
	\begin{equation*}
	p_\rho(x|\Pi)=\tr{\rho\Pi_x}. 
	\end{equation*}
	$\Pi$ is often called a {\it positive operator-valued measure} (POVM) in the literature. 
We denote the set of all possible POVMs on $\cH$ by $\cM(\cH)$. \\[1ex]
\noindent
Q4) \underline{Channels}. 
	A {\it quantum channel} (also known as a {\it quantum process}) $\cT$ is a linear map from the input quantum state space $\sofh$ to the output state space ${\cal S}({\cal H}')$. We only consider cases where $\cH'=\cH$. 
	Axiomatically, a channel is defined as a completely positive and trace-preserving map \cite{NC,petz}. 
	A convenient representation of a quantum channel is the {\it Kraus representation}, defined as
	\begin{equation*}
	\cT(\rho)=\sum_{k=1}^K E_k \rho E_k^{\dagger},
	\end{equation*}
	where the Kraus operators $E_k\in\bbc^{d\times d}$ satisfy the trace-preserving condition: $\sum_{k=1}^KE_k^{\dagger} E_k=I_d$.

\subsection{Formulation of the problem}
	We can now formulate the problem of quantum channel parameter estimation in the framework of \odoe{}. 
	We start with a family of $n$-parameter quantum channels 
	\begin{equation*}
	M^Q=\{\cT_{\param{}}\,|\,\param{}\in\Param\subset\bbr^n\},
	\end{equation*}
	assuming that $\param{}\mapsto\cT_{\param{}}$ is one-to-one and smooth mapping. 
	The design $e$ is a set of input quantum states $\rho\in\sofh$ and a POVM $\Pi\in\cM(\cH)$ on the output quantum state $\cT_{\param{}}(\rho)$, i.e., $e=(\rho,\Pi)$. The design space is $\cE=\sofh\times\cM(\cH)$. 
	The model function $ f $ is given by Born's rule and the resulting probability distributions are
	\begin{equation*}
	p_{\param{}}(x|e)=\tr{\cT_{\param{}}(\rho) \Pi_x}, 
	\end{equation*}
	for a given quantum channel and a chosen design $e=(\rho,\Pi)$. 
	Thus, the statistical model is 
	\begin{equation*}
	M(e)=\{p_{\param{}}(\cdot|e)\,|\,\param{}\in\Param\}. 
	\end{equation*}
	
	We wish to find an optimal design $e_*=e_*(m_*)=(\v{\nu},\v{e})\in\cP(m_*)\times\cE^{m_*}$ 
	that minimizes a properly chosen optimality criterion, i.e., 
	\begin{align}
	e_*(m)&=\arg\min_{e(m)}\Psi(J_{\param{}}[e(m)] ),\nonumber\\
	m_*&=\arg\min_{m\in\bbn}\Psi\left(J_{\param{}}[e_*(m)] \right). 
	\end{align} 
Solving the optimization problem, however, can be difficult because the design is composed of two distinct parts: a state $\rho$ and a measurement $\Pi$.
	
	This difficulty can be partially assuaged by introducing the quantum extension of the Cram\'er-Rao bound \cite{helstrom,holevo,petz},
	\begin{equation}\label{eq:qm.crb}
	V_{\param{}}[\hparam{}|e]\ge \Big( J_{\param{}}[e]\Big)^{-1} \ge \Big( J_{\param{}}^{QM}[\rho] \Big)^{-1}
	\end{equation}
Here, $J_{\param{}}^{QM}[\rho]$ is the quantum Fisher information (QFI), which depends only on the input state $\rho$.
Just as its classical counterpart is a measure of how much information about a parameter can be extracted from a statistical model, the quantum Fisher information is a measure of how much information about parameters $ \param{} $ can be extracted from a quantum state. 
	We consider only the Symmetric Logarithmic Derivative (SLD) QFI $J_{\param{}}^{QM}=\qfi{\param{}} $, defined as
	\begin{equation}\label{eq:qfi.def}
	[\qfi{\param{}}]_{ij} = \tfrac{1}{2}\Tr{\cT_{\param{}}(\rho)(\mathcal{L}_{\param{},i}\mathcal{L}_{\param{},j} + \mathcal{L}_{\param{},j} \mathcal{L}_{\param{},i})},
	\end{equation}
	where the quantum score functions $ \mathcal{L}_{\param{},i} $ are solutions to the equation
	\begin{equation} \label{eq:sld.score}
	\frac{\partial \cT_{\param{}}(\rho)}{\partial \param{i}} =\frac12 \mathcal{L}_{\param{},i} \cT_{\param{}}(\rho) + \frac12\cT_{\param{}}(\rho) \mathcal{L}_{\param{},i}.
	\end{equation}
	
	The matrix inequality in Eq.~(\ref{eq:qm.crb}) follows from the monotonicity of the SLD QFI under further action of a quantum channel (in this case, that of the POVM, considered as a quantum channel) \cite{petz96}. 
	The second inequality in Eq.~\eqref{eq:qm.crb} cannot be, in general, saturated, but this inequality is useful 
	for deriving a bound for a given optimality function as 
	\begin{equation}\label{eq:qfi.bound}
	\Psi( J_{\param{}}[e(m)])\ge \Psi(J_{\param{}}^{QM}[(\v{\nu},\v{\rho})])\ge\Psi(J_{\param{}}^{QM}[(\v{\nu}_*,\v{\rho}_*)]), 
	\end{equation}
	where 
	\begin{equation}
	J_{\param{}}^{QM}[(\v{\nu},\v{\rho})]\equiv \sum_{i=1}^m\nu_iJ_{\param{}}^{QM}[\rho_i]
	\end{equation}
	and $(\v{\nu}_*,\v{\rho}_*)=\arg\min_{\v{\nu},\v{\rho}} \Psi(J_{\param{}}^{QM}[(\v{\nu},\v{\rho})])$ optimizes the given 
	optimality criterion. Since this optimization involves input states $\v{\rho}=(\rho_1,\dots,\rho_m)$ only, it is much easier to handle. 
	It is clear that all the previously reviewed methods of finding an optimal design using the Fisher information are also applicable to this optimization problem about the quantum Fisher information.
Note that if the lower bound set by the QFI is saturated by the classical Fisher information and $J_{\param{}}[e] = \qfi{\param{}}[\rho] $, 
the CR bound is then likewise saturated and the Fisher information is equal to the MSE matrix \cite{young,nagaoka87,bc94}. 
	
As in the classical case, the above optimization problem generally yields a \emph{local} optimal design. The solution then depends on the unknown parameter $\param{}$ in general. We stress again that $\param{}$-dependent optimal estimation strategies, in particular, $\param{}$-dependent optimal measurements, are generic in the theory of optimal DoE. In the context of quantum state estimation problems, a number of authors proposed and analyzed adaptive methods to implement such $\param{}$-dependent POVMs; see for example Refs.~\cite{nagaoka89-2,HM98,BNG00,fujiwara06,stm12}. Experimental realization of these adaptive estimation methods are also an active subject over the last decade; see Refs.~\cite{oioyift12,mrdfbks13,ksrhhk13,hzxlg16,ooyft17} and also review article \cite{zlwjn17} on the subject.

\subsection{Discussions and extensions}
Several remarks are in order about the extension of \odoe{} to the quantum setting. First, usually not all input states are realizable in experiments and $\rho$ can only come from a subset of $\sofh$, say ${\cal S}_0$. 
Similar practical constraints also often apply to the measurement space, and one can consider only a subset of all measurements, $\cM_0\subset\cM(\cH)$. 
A common restriction is to take $ \cM_0 $ as the set of projective measurements, or projection-valued measures (PVMs), denoted by $\cM_{PVM}$. 
	The design space is then $\cE={\cal S}_0\times \cM_{PVM}$. 
	
We also list three variants of possible design spaces that may arise from experimental constraints. 
\begin{enumerate}
\item If the input state is fixed such that ${\cal S}_0=\{\rho_0\}$, the problem is reduced to that of quantum state estimation. In this case, we optimize over only the POVM $\v{e}=(\Pi(i))$ and relative frequencies $\v{\nu}=(\nu_i)$.

\item When the measurement $\Pi$ is fixed, on the other hand, we see that the problem becomes one of finding the best set of input states and relative frequencies $\v{\nu}$. One of us (J. Suzuki) has already reported on this problem for the channel-parameter estimation problem in classical information theory \cite{js16sita}. A general formula for the optimal design for a binary-input two-parameter case is given in the next subsection. 
	
	Let us briefly go over this problem. We let $J_{\param{}}[\rho]$ be the Fisher information matrix for an input state $\rho$ with fixed measurement $\Pi$. 
	The Fisher information matrix for the design $e(m)=(\v{\nu},\v{e})$, where
	$\v{e}=(\rho_1,\rho_2,\dots,\rho_m)$ is now a set of input states, is
	\begin{equation}
	J_{\param{}}[e(m)]=\sum_i \nu_iJ_{\param{}}[\rho_i]. 
	\end{equation}
Since the Fisher information matrix is convex with respect to the input state, i.e.,
	\begin{equation}
\hspace*{1cm}	J_{\param{}}[p \rho_1 + (1 - p)\rho_2]\le p J_{\param{}}[\rho_1]+(1-p)J_{\param{}}[\rho_2], 
	\end{equation}
	$\forall p \in[0,1]$, the optimal input states are pure states. 
	This point is important when dealing with the general optimization case. 
	Since this statement is true for any POVM, optimal input states are always pure states \cite{fujiwara01}. 
        In other words, we can always restrict to optimal pure input states and then optimize over the POVM.
	
\item If both the input set ${\cal S}_0\subset\sofh$ and the POVM set $\cM_0\subset\cM(\cH)$ are fixed, we optimize over only the relative frequencies $\v{\nu}=(\nu_1,\nu_2,\dots,\nu_m)$. 
	We will use the standard process tomography setting as an example, where one adopts the design $e_i=(\psi_i,\Pi_i)$. 
	Here, $\psi_i$ are pure states and $\Pi_i$ are the corresponding PVMs, such that $\{e_i\}$ comprises an informationally complete estimation strategy for the quantum channel space. 
	This class of optimal design problems was discussed in Refs.~\cite{kwr04,nunn10}. 
\end{enumerate}
	
	A convex structure for the design space $\cE$ can be introduced as follows. 
	For the input states $\rho_1,\rho_2$, the convex sum of two states is defined as $\rho_p=p\rho_1+(1-p)\rho_2$, for $p\in[0,1]$, which is still in the set $\sofh$. 
	For two measurements $\Pi(1)=\{\Pi_1,\dots,\Pi_{k_1}\}$ and $\Pi(2)=\{\Pi'_1,\dots,\Pi'_{k_2}\}$, 
	we can define a convex sum as 
	$\Pi_p=p\Pi(1)\bigcup(1-p)\Pi(2)=\{p\Pi_1,\dots,p\Pi_{k1},(1-p)\Pi'_1,\dots,(1-p)\Pi'_{k_2}\}$. 
	Statistically, this convex sum is equivalent to performing measurement $\Pi(1)$ 
	with probability $p$ and measurement $\Pi(2)$ with probability $1-p$. 
	Such a measurement is called a {\it randomized measurement} since it can be realized with (pseudo-)random numbers \cite{dpp05,fujiwara06,yamagata11}. 
	Here we see that the theory of optimal design of experiments unifies previously studied optimization problems in a systematic way.
	
	Lastly, we briefly mention possible extensions of the above estimation strategy to other estimation settings. 
	There are several distinct strategies for utilizing quantum resources, such as entangled states, ancilla states, or joint measurements on output states for quantum estimation problems. 
	It has been known in the literature that these extended estimation strategies can lower the estimation errors in general. 
	We have so far not mentioned these methods, but they can also be formulated as optimal experimental design problems. 
	
	As an example, let us consider an ancillary assisted estimation strategy. 
	Let $\cH_A$ be the Hilbert space of the ancilla states and $id_{A}$ be the identity map on it. 
	The family of quantum channels to be estimated is expressed as $\{\cT_{\param{}}\otimes id_A|\param{}\in\Param\}$ and the input state space is extended to ${\cal S}(\cH\otimes\cH_A)$. 
	Likewise, the measurement space can be extended. 
	Then, the optimization problem takes the same form as before except that the design space is extended.

\subsection{Analytical results}
A closed-form expression for an optimal design cannot usually be obtained analytically except in very special cases. 
In the following, we briefly discuss some of these special cases.

\subsubsection{L\"owner optimal design} \label{sec-loevner}

	As mentioned before, the existence of the L\"owner optimal design is a special case where a closed-form expression can be derived. 
	Suppose there exists a design $e_*$ that is L\"owner optimal, and its expression is obtained. 
	We show below that mixed strategies do not give any advantage over the i.~i.~d.~strategy for most popular optimality criteria.
	
	Consider the i.i.d strategy, using $e_*$ repeatedly. This design is also L\"owner optimal for any design $e(m)$ since any mixed strategy $e_p=\big((p, 1-p), (e_1,e_2)\big)$ for $m=2$ 
	with $e_1,e_2\in\cE$ obeys the inequality
	\begin{align*}
	J_{\param{}}[e_p]&=p J_{\param{}}[e_1]+(1-p)J_{\param{}}[e_2]\\
	&\le p J_{\param{}}[e_*]+(1-p)J_{\param{}}[e_*]=J_{\param{}}[e_*].
	\end{align*}
	Therefore, $e_*(2)=\big((p, 1-p), (e_*,e_*)\big)$. 
	We can repeat this argument to show that $e_*(m)$ has a similar structure, and conclude that an optimal design is the i.~i.~d.~strategy. 
	
	Next, remember that when a L\"owner optimal design exists, it is also optimal for other optimality criteria. 
	Consider an optimality function $\Psi$ satisfying the isotonicity property discussed in Appendix Sec.~\ref{sec-app_supp2}. 
This, together with the argument that the i.~i.~d.~strategy is optimal in the L{\"o}wner optimal case, tells us that any mixed strategy cannot further minimize the function $\Psi(J_{\param{}}[e(m)])$. 
	
\subsubsection{Single-parameter family of quantum channels}\label{sec-1para}
	When considering a single-parameter family of quantum channels, we can find an optimal solution analytically 
in the language of the theory of \odoe{} (see also Refs.~\cite{sm06,js16pra}.)
	Let $M^Q=\{\cT_{\param{}}|\param{}\in\Param\subset\bbr\}$ be a one-real-parameter family of quantum channels. The design space is $\cE = \sofh\times \cM(\cH)$.
	Eq.~(\ref{eq:qm.crb}) then gives 
	\begin{align}
	\label{1paraopt2}J_{\param{}}[(\rho,\Pi)]&\le J^{\sld}_{\param{}}[\rho]\\ 
	&\le J^{\sld}_{\param{}}[\rho_*]. \label{1paraopt}
	\end{align}
	An optimal measurement that attains the first equality [Eq.~\eqref{1paraopt2}] is known \cite{young,nagaoka87,bc94}. 
	The second inequality [Eq.~\eqref{1paraopt}] follows from the maximization of the SLD quantum Fisher information over all input states,
	\begin{equation}
	\rho_*=\arg\min_{\rho:\sofh}J^{\sld}_{\param{}}[\rho].
	\end{equation}
	Note that the optimizer $\rho_*$ is not unique in general and it can always be a pure state as argued earlier.  
	Hence, we can bound all possible classical Fisher information by the optimal one in Eq.~\eqref{1paraopt}. This then must be the L\"owner optimal design and optimal among all possible designs including the mixed strategy.

\subsubsection{Two-parameter binary-design problem}\label{sec:twoBin}
	Let us consider a generic two-parameter binary-design problem, where $M^Q=\{\cT_{\param{}}\,|\,\param{}=(\param{1},\param{2})\in\Param\}$ and the design space has only two elements $\cE=\{e_1,e_2\}$. 
	In the quantum setting, this is equivalent to setting ${\cal S}_0=\{\rho_1,\rho_2\}$ and fixing the POVMs for each output state $\cT_{\param{}}(\rho_i)$ as $\Pi_i$ ($i=1,2$). 
	We assume that the corresponding statistical model 
	\begin{equation}
	M(e_i)=\{p_{\param{}}(\cdot|e_i)\,|\,\param{}\in\Param\}, 
	\end{equation}
	is regular. 
	
We first analyze the conditions for the existence of the L\"owner optimal design here. Let us introduce some notation for our convenience. Let $J_{i}$ be the Fisher information matrices for the $i$th model $M(e_i)$, i.e., $J_1=J_{\param{}}[e_1]$ and $J_2=J_{\param{}}[e_2]$, and define
\begin{align*}
&T_1=\tr{J_1},\quad T_2=\tr{J_2},\\
&D_1=\Det{J_1},\ D_2=\Det{J_2},\ D_{\pm}=\Det{J_1\pm J_2}.  
\end{align*}
We assume that $J_1, J_2$ are positive definite, i.e., the designs $e_1,e_2$ are regular. A L\"owner optimal design exists when a matrix ordering is possible, i.e., $J_1\ge J_2$ (if $J_1\le J_2$, we swap the labeling $1\leftrightarrow 2$). 

For a symmetric $2\times2$ matrix $A$, which is not equal to the zero matrix, the positive semidefinite relation 
$A \ge 0$ fails to hold if and only if $A$ has two distinct eigenvalues with opposite signs. 
This is equivalent to $\Det{A}<0$. 
Therefore, the following case is the generic one to be analyzed,
\begin{equation}\label{2paracond}
D_-=\Det{J_1-J_2} <0,
\end{equation}
which, if satisfied, indicates that there is no L\"owner optimal design.
	
The $A$- and $D$-optimal designs for this two-parameter binary-design problem can be found analytically.

\smallskip
\noindent\underline{$D$-optimality}. 
	For two given designs, the optimization problem is equivalent to finding the optimal relative frequency $\v{\nu}=(\nu_1,\nu_2)$ such that $\Det{\nu_1J_1+\nu_2J_2}$ is maximized. When the optimal $\v{\nu}_*$ is located at extremal points, i.e., either $(1,0)$ or $(0,1)$, 
	an optimal design is the i.~i.~d.~strategy. This is because using any mixed strategy cannot further 
	maximize the function $\Det{\nu_1J_1+\nu_2J_2}$. 
	
We parametrize $\nu_1=\frac{1}{2}(1+\lambda)$ and $\nu_2=\frac{1}{2}(1-\lambda)$, with $\lambda\in[-1,1]$, and define the function 
	\begin{align*}
	\gamma_{\param{}}(\lambda)&=4\Det{\nu_1J_1+\nu_2J_2}\\
	&=\Det{J_1+J_-+\lambda (J_1-J_2)}\\
	&=D_-\lambda^2+2 (D_1-D_2)\lambda +D_+.
	\end{align*}
	Since we are considering the case where there is no L\"owner optimal design, condition \eqref{2paracond} needs to be imposed. 
	The optimal $D$-design is then found by maximizing the quadratic function $ \gamma_{\param{}}(\lambda) $, yielding 
	\begin{equation}
	\max_{\lambda\in[-1,1]}\gamma_{\param{}}(\lambda)=
	\begin{cases}
	\gamma_{\param{}}(\lambda^*)\qquad (\textrm{if }|D_1-D_2|<-D_-),\\[1ex]
	\max\{ \gamma_{\param{}}(1),\gamma_{\param{}}(-1)\}\quad (\textrm{otherwise}),
	\end{cases}
	\end{equation}
	with $\lambda^*=-(D_1-D_2)/D_-$. 
	The optimal design is then $\v{\nu}_*=\big(\frac{1}{2}(1+\lambda^*),\frac{1}{2}(1-\lambda^*)\big)$ when $|D_1-D_2|<-D_-$ is satisfied; otherwise, the optimal design is extremal, $\v{\nu}=(1,0)$ or $(0,1)$, depending on $\arg\max\{ \gamma_{\param{}}(1),\gamma_{\param{}}(-1)\}$. 
	In the latter case, an optimal design is the i.~i.~d.~strategy as mentioned earlier. 
	However, the relation $ \gamma_{\param{}}(1)\ge\gamma_{\param{}}(-1)\Leftrightarrow D_1\ge D_2\Leftrightarrow \Det{J_{\param{}}[e_1]}\ge\Det{J_{\param{}}[e_2]}$ indicates that an optimal design at $\theta$ 
	may depend on the unknown value of $\theta$ in general. This is the typical behavior of local optimal designs.
	
Note that the case where both $e_1$ and $e_2$ are singular designs can also be treated similarly. In that case,  $D_1=D_2=0$, which simplifies $ \gamma_{\param{}}(\lambda) $, giving $\gamma_{\param{}}(\lambda)=D_-\lambda^2+D_+$. The optimal design in this case is then $\v{\nu}_*=(1/2,1/2)$. 

\smallskip
\noindent\underline{$A$-optimality}. 
	We now consider $A$-optimal designs with a weight matrix $W>0$. 
	We now define the function (of $\lambda$) $ \gamma_{\param{}}[W](\lambda) $, dependent on both $W$ and $\param{}$, as 
	\begin{equation} \label{eq:gamfun.defn}
	\gamma_{\param{}}[W](\lambda)=\Tr{W{\left[\tfrac{1}{2}(1+\lambda)J_1+\tfrac{1}{2}(1-\lambda)J_2\right]}^{-1}}.
	\end{equation}
	We can set a lower bound for the $A$-optimal design as
	\begin{equation}
	\Psi_A(e(2))\ge \gamma_{\param{}}^*[W]= \min_{\lambda\in[-1,1]}\gamma_{\param{}}[W](\lambda). 
	\end{equation}
	 The following result is our contribution: For a two-parameter binary-design problem, when condition \eqref{2paracond} holds, the bound for the $A$-optimal design is given by (from straightforward, though lengthy, calculations)
	\be \label{2pararesult}
	\gamma^*_{\param{}}[W]=
	\begin{cases}
		\ds \frac{D_1\gamma_{\param{}}(1)}{2\max\{D_1,D_2\}}+ \frac{D_2\gamma_{\param{}}(-1)}{2\max\{D_1,D_2\}}\\
		\qquad\Big(\mathrm{if }\ D_1 \gamma_{\param{}}(1)=D_2\gamma_{\param{}}(-1),\\
		\hspace{2cm} \textrm{and }|D_1-D_2|-|D_-|>0\Big),\\[1ex]
		\ds\frac{2D_-D_1\gamma_{\param{}}(1)+2D_-D_2\gamma_{\param{}}(-1)}{D_+D_- -(D_1-D_2)^2}\\
		\qquad\Big(\mathrm{if}\ D_1 \gamma_{\param{}}(1)=D_2\gamma_{\param{}}(-1),\\
		\hspace{2cm}\textrm{and } |D_1-D_2|-|D_-|\le0\Big),\\[1ex]
		\gamma_{\param{}}(\lambda^*)\quad\Big(\mathrm{if }\ D_1 \gamma_{\param{}}(1)\neq D_2\gamma_{\param{}}(-1),\\
		\hspace{2cm}\gamma_{\param{}}(\lambda_\pm)>0,\,\textrm{and } |\lambda_*|\le1\Big),\\[1ex]
		\min\{\gamma_{\param{}}(1),\gamma_{\param{}}(-1) \}\quad (\mathrm{otherwise}).
	\end{cases}
	\ee
	Here, $\lambda_\pm (\lambda_+\ge\lambda_-)$ are roots of the quadratic equation
	\be
	D_-\lambda^2+2(D_1-D_2)\lambda+D_+=0,
	\ee
	and $\lambda_*$, characterizing the optimal input, is given by
	\begin{equation} \label{eq:opt.lambda}
	\lambda_*=\frac{\sqrt{\gamma_{\param{}}(-1)}\lambda_++\sqrt{\gamma_{\param{}}(1)}\lambda_- }{\sqrt{\gamma_{\param{}}(1)}+\sqrt{\gamma_{\param{}}(-1)}}. 
	\end{equation}
Note that, in Eq.~\eqref{2pararesult}, the first two cases are special ones, since $D_1 \gamma_{\param{}}(1)=D_2\gamma_{\param{}}(-1)$ is equivalent to $\tr{WJ_1^{-1}}=\tr{WJ_2^{-1}}$. This is satisfied for a specific choice of the weight matrix. 
	The third case is when the mixed estimation strategy brings the estimation error below that of the  i.~i.~d.~strategy. The last case is when an optimal design is located at extremal points.

\section{Examples}\label{sec4}
In this section, we analyze families of qubit channels as examples to illustrate our findings and point out special features. As a benchmark, we compare the \odoe{} strategies with that of the a simple, and commonly used, quantum process tomography scheme built upon the Pauli operators $\{\sigma_i\}_{i=1}^3$ (we denote this as Pauli-QPT): For each $i=1,2,3$, we send, as input to the channel, the eigenstates of $\sigma_i$, $\rho_\pm(i)$ with $\pm1$ eigenvalues, and then perform the projective measurement $\Pi(i)=\{ (I_d\pm\sigma_i)/2\}$ on the output state with the uniform probability. For the channels discussed in this section, it happens that 
$\rho_\pm(i)$ for each $i$ gives the exact same Fisher information for the projective measurement. Thus, we only consider 
the $+1$ eigenstates with equal probability $1/3$.

\subsection{Linear scaling channel} \label{sec4-1}
	Let us start with the simplest example in which we consider a three-parameter family of qubit channels specified as
	\begin{equation} \label{eq:scaling.chan}
	\cT_{\param{}}(\rho) \leftrightarrow \vec{s}_{\param{}}= \begin{pmatrix}
	\param{1}s_1 \\ \param{2}s_2 \\ \param{3}s_3
	\end{pmatrix}.
	\end{equation}
Here, $\vec s_{\param{}}$ is the Bloch vector of the output state $\cT_{\param{}}(\rho)$, $\vec{s}=(s_1,s_2,s_3)^\mathrm{T}$ is that of the input state $\rho$, and $\theta\in\Theta=\{ \theta\in\bbr^3| \sum_{i=1}^3 |\theta_i|^2\le 1\}$. Each member of this family of channels linearly scales the Bloch components of the input state. 
Some authors refer to this channel as a generalized Pauli channel \cite{po09,bh10,bh11}. 
Note that, following the procedure in Appendix Sec.~\ref{sec-app_lowner}, we can 
	show that there is no L\"owner optimal design. 
	
For this example, the SLD quantum Fisher information matrix for the input state $ \vec{s}=(s_1, s_2, s_3)^\mathrm{T}$ can be written as
	\begin{equation} \label{eq:scaling.qfi}
	\qfi{\param{}}[\rho] = D(s)^{1/2}\left[ I + \frac{\vec{s}_{\param{}} \vec{s}_{\param{}}^{\hspace{0.6mm}T} }{1 - \vec{s}_{\param{}}^{\hspace{0.6mm}T}\vec{s}_{\param{}}} \right]D(s)^{1/2}
	\end{equation}
	where $ D(s) $ is the positive matrix $\textrm{diag}(s_1^2,s_2^2, s_3^2) $.
	The convex structure of the problem means that an optimal design is composed of extremal points of the space of SLD quantum Fisher information matrices for all possible input states. These extremal points are rank-1 matrices.
	From Eq.~(\ref{eq:scaling.qfi}), it is clear that a rank-1 $ \qfi{\param{}} $ must have a rank-1 $ D(s) $. 
	This corresponds to having $ \vec{s} $ in the 1, 2, or 3, direction, the Bloch vectors of the eigenstates of the Pauli operators. 
	
	For such input states, projective measurements allow for designs that saturate the lower bound (as the Fisher information matrix will be singular), set by the SLD quantum Fisher information, i.e., $ J_{\param{}}[e_i] = \qfi{\param{}}[e_i] $, for $ i = 1, 2, 3 $. Such projective measurements correspond to measurement operators defined by $\Pi_{i,\pm} = \frac{1}{2}( I_d \pm \sigma_i)$; the eigenstate and the projective Pauli measurement together make the design $ e_i $.
	
An optimal design is hence composed as a mixture of the three Pauli settings $\bigl(e_i=(\rho(i),\Pi(i))\bigr)$ with relative frequencies $\v{\nu}=(\nu_1,\nu_2,\nu_3)$, giving the Fisher information matrix
\begin{equation}
	J_{\param{}}[e(3)]=\mathrm{diag.}{\left(\frac{\nu_1}{1-\param{1}^2},\frac{\nu_2}{1-\param{2}^2},\frac{\nu_3}{1-\param{3}^2}\right)}. 
\end{equation}
	By optimizing the $\v{\nu}$ degree of freedom, one can find the best estimation strategy, according to the desired optimality criterion. Observe that the Pauli-QPT design $e_{PT}$ is the case where $\v{\nu}=(1/3,1/3,1/3)$, so that
	\begin{equation}
	J_{\param{}}[e_{PT}]=\frac13 \mathrm{diag.}{\left(\frac{1}{1-\param{1}^2},\frac{1}{1-\param{2}^2},\frac{1}{1-\param{3}^2}\right)}. 
	\end{equation}
	
As an example, let us find the $\gamma$-optimal design for $\gamma>0$. 
The application of Jensen's inequality and the convexity of $x^{1/(1+\gamma)}$ for $\gamma>0$ give
\begin{align*}
		&\min_{p\in\cP(m)}\sum_{i=1}^m{\left(\frac{a_i}{p_i}\right)}^\gamma= \biggl(\sum_{i=1}^ma_i^{\frac{\gamma}{1+\gamma}}\biggr)^{1+\gamma} ,\\[1ex]
\textrm{and}\quad 		&p_*=\arg\min_{p\in\cP(m)}\sum_{i=1}^m{\left(\frac{a_i}{p_i}\right)}^\gamma= a_i^{\frac{\gamma}{1+\gamma}}/\sum_j a_j^{\frac{\gamma}{1+\gamma}} ,
		\end{align*} 
for any a positive $m$-dimensional vector $a=(a_i)\in\bbr_+^m$. 
Here, $\cP(m)$ denotes the set of $m$-event (positive) probability distributions. 
This immediately solves the optimization problem at hand and yields the $\gamma$-optimal solution:
	\begin{equation*}
	\min_{\nu\in\cP(3)} \left\{\frac13\Tr{J_{\param{}}[e(3)]^{-\gamma}}  \right\}^{1/\gamma}
	=\Big[\sum_i (\frac{1-\param{i}^2}{3})^{\frac{\gamma}{1+\gamma}}\Big]^{\frac{1+\gamma}{\gamma}}.
	\end{equation*}
	As an example,  the $A$-optimal design ($\gamma=1$) $e_*(3)$ has
	\begin{equation*}
	\nu_i=\frac{\sqrt{1-\param{i}^2}}{\sum_{j} \sqrt{1-\param{j}^2}}.
	\end{equation*}
	Observe that $\Tr{J_{\param{}}[e_{PT}]^{-1}}\ge\Tr{J_{\param{}}[e_{*}(3)]^{-1}}$ always holds, with equality 
	if and only if $\param{1}=\param{2}=\param{3}$, the case of an isotropic channel.

\subsection{Pauli channel}
	The {\it Pauli channel} for the qubit is defined by 
	\begin{equation}\label{eq:PauliChannel}
	\cT_{\param{}}(\rho)=(1-\sum_{i}\param{i}) \rho+\sum_{i=1,2,3}\param{i}\sigma_i\rho\sigma_i, 
	\end{equation}
	where the channel parameter $\param{}=(\param{1},\param{2},\param{3})$ are all positive 
	and the sum of them is less than one ($1-\sum_{i}\param{i}>0$). 
	In the Bloch vector representation, a state $\vec{s}$ is transformed as 
	\begin{equation*}
	\cT_{\param{}}:\, \vec{s}\mapsto \vec{s}_{\param{}}= (\xi_1(\param{})s_1,\xi_2(\param{})s_2,\xi_3(\param{})s_3) ^\mathrm{T},
	\end{equation*}
	where $\xi_i(\param{})=1+2\param{i}-2\sum_{j}\param{j}$. Thus, the Pauli channel can be regarded 
	as a different coordinate system representation (under an affine coordinate transformation) of the linear scaling channel. Therefore, a L\"owner optimal design for this Pauli channel problem cannot exist, as was the case for the linear scaling channel. 
	
It is nevertheless interesting to see how the optimal design depends on the parameterization of the channel family. Following the same reasoning as before, an optimal design $e_*(3)$ is again a mixture of the Pauli settings $e_i=(\rho(i),\Pi(i))$ for $i=1,2,3$, with relative frequencies $\v{\nu}=(\nu_i)$. Its Fisher information matrix is given by
	\begin{equation}
	J_{\param{}}[e(3)]=\sum_{i=1,2,3}\frac{4\nu_i}{1-(\xi_i)^2}\vec{u}_i \vec{u}_i^\mathrm{T},
	\end{equation}
	with $\vec{u}_1=(0,1,1)^\mathrm{T}, \vec{u}_2=(1,0,1)^\mathrm{T}$, and $\vec{u}_3=(1,1,0)^\mathrm{T}$, three non-orthogonal vectors.
	
	Unlike the linear scaling example, the analytical formula for $\gamma$-optimality here cannot be expressed as a closed-form solution in general. The $A$-optimal ($\gamma=1$) solution, however, can be found as
	\begin{equation*}
	\min_{\v{\nu}}\Tr{J_{\param{}}[e_{3}]^{-1}}=\frac{3}{16}{\biggl[\sum_i\sqrt{1-(\xi_i)^2}\biggr]}^2.
	\end{equation*}
	The corresponding $A$-optimal design is given by the mixture of the Pauli settings with the optimal relative frequencies 
	\begin{equation*}
	\nu_i=\frac{\sqrt{1-(\xi_i)^2}}{\sum_{j} \sqrt{1-(\xi_j)^2}}.
	\end{equation*}
	The $D$-optimal ($\gamma\rightarrow 0$) solution, on the other hand, is given by
	\begin{align*}
	\min_{\v{\nu}}\Det{J_{\param{}}[e_{3}]^{-1}}
	&=\max_{\v{\nu}}2^{-8}\prod_i\frac{\nu_i}{1-(\xi_i)^2}\\
	&= 2^{-8}3^{-3} \prod_i\frac{1}{1-(\xi_i)^2}, 
	\end{align*}
	This $D$-optimal design coincides with the Pauli-QPT design, with $\v{\nu}_*=(1/3,1/3,1/3)$. This simple example shows that different optimality criteria result in different optimal designs.

\subsection{Detecting noise asymmetry}\label{sec5-3}

We now turn to an example where a nuisance parameter arises naturally. We consider a two-parameter family of Pauli channels as in Eq.~\eqref{eq:PauliChannel}, all with $\param{3}=0$. It is convenient to use a different parameterization to describe the family,
\begin{equation} \label{eq:noise.params}
\begin{aligned}
\vparam{1} &\equiv \param{1} - \param{2}, \\
\vparam{2} &\equiv 1-(\param{1} + \param{2}).	
\end{aligned}
\end{equation}
Here, $\vparam{1}\in[-1,1]$ and $\vparam{2}\in[0,1]$, with $\vparam{2}\leq 1-|\vparam{1}|$ \cite{footnote2}. 
$\vparam{1}$ is the \emph{asymmetry} of the channel, characterizing the imbalance between the strength of the $\sigma_1$ and $\sigma_2$ Kraus operators; $1-\vparam{2}=\param{1}+\param{2}$ describes the deviation of the channel from the identity operation. Denoting the channel as $\cT_{\vparam{}}$, its action on the state Bloch vector is
\begin{equation}
\cT_{\vparam{}}(\rho)\leftrightarrow \vec{s}_{\vparam{}}= {\left(
	\begin{array}{c}
	(\vparam{1}+\vparam{2})s_1\\
	(-\vparam{1}+\vparam{2})s_2\\
	(2\vparam{2}-1)s_3
	\end{array}
	\right)}.
\end{equation}

Viewing the Pauli channel as noise, $1-\vparam{2}$ is the noise strength, and $\vparam{1}$ is the noise asymmetry. $ \vparam{1} $ is a practically useful quantity in the control of noise in quantum information processing.
Noise with a large asymmetry can be mitigated more efficiently by first reducing the noise asymmetry with a small error-correcting code, before a more resource-intensive code that does not pay attention to the asymmetry is used to reduce the overall noise strength.
Within this context, we are interested in estimating the asymmetry $\vparam{1}$; $\vparam{2}$ is treated as a nuisance parameter. The task is to discover the optimal design for estimating $\vparam{1}$.

\subsubsection{Optimal design problem}\label{sec5-3-1}
Following the notation from Sec.~\ref{sec2-3}, our two-parameter problem $\vparam{}=(\vparam{1},\vparam{2})$ is split into the parameter of interest, $\vparam{I}= \vparam{1}$, and the nuisance parameter $\vparam{N}=\vparam{2}$. 
The presence of a nuisance parameter complicates the formal solution of the optimal design problem, compared to a full channel characterization. 

For a mixed strategy with $m$ partitions of the total input states, Carath\'eodory's theorem mentioned in Sec.~\ref{sec2-5} ensures that an optimal design can be found for $m\le 7$. 
For a given $m$, we simplify the search for an optimal design $e_\mathrm{opt}=(\rho_\mathrm{opt},\Pi_\mathrm{opt})$ with a two-step approach.
We first optimize the SLD Fisher information over $\rho$, which fixes $\rho_\mathrm{opt}$, and then optimize the classical Fisher information over $\Pi$ for the chosen $\rho_\mathrm{opt}$. 
Once we have found the optimal designs for each $ m $, we then compare them to find the true optimal design for estimating the noise asymmetry. 
Focusing on closed analytical forms for the optimal design, we will work up to $m=2$ only. For $m=3$, we analyze the specific case of Pauli-QPT, and compare its performance with that of the $m=2$ optimal design, as an indication of how much benefit one might expect from increasing $m$.

We make one further simplifying assumption, that $s_3=0$ for $\rho_\mathrm{opt}$. This is reasonable, given that the transformation of $s_3$ under $\cT_{\vparam{}}$ does not involve the parameter of interest $\vparam{1}$. This significantly simplifies our analysis here.
With $s_3=0$, straightforward algebra gives 
\begin{equation}
\qfi{\vparam{}}=\frac{\left(s_1^2\vec{v}_1\vec{v}_1^\mathrm{T}+s_2^2\vec{v}_2\vec{v}_2^\mathrm{T}-2s_1^2s_2^2\vec{u}\vec{u}^\mathrm{T}\right)}{g(s_1,s_2)}\,,
\end{equation}
where ${\vec{v}_1=(1,1)^\mathrm{T}/\sqrt{2}}$, ${\vec{v}_2=(1,-1)^\mathrm{T}/\sqrt{2}}$, ${\vec{u}=(-\vparam{2},\vparam{1})^\mathrm{T}}$, 
and ${g(s_1,s_2)= \frac{1}{2}(1-|\vec{s}_{\vparam{}}|^2)}$. 
This SLD quantum Fisher information has determinant
\begin{equation}
\mathrm{Det}(\qfi{\vparam{}}[\rho])=\frac{2s_1^2s_2^2}{g(s_1,s_2)},
\end{equation}
which vanishes whenever $s_1s_2=0$. 

For the $m=1$ case, the local optimal design can be found by looking for $e(1)$ that satisfies
\begin{equation}\label{eq:loc_sing_opt}
\min_{s_1,s_2:\,\sum_is_i^2=1} {\left(\qfi{\vparam{}}[e(1)]^{-1}\right)}_{11}=\min \{ f_1^2\,,\,f_2^2  \} ,
\end{equation}
where we define ${ f_{1,2} = \frac{1}{2}\sqrt{1-(\vparam{1}\pm\vparam{2})^2}}$, and the inverse is the generalized inverse
The solution is the singular design $(\rho_\pm(1),\Pi(1))$ 
if $f_1^2< f_2^2 \Leftrightarrow \vparam{1}>0$; it is $(\rho_\pm(2),\Pi(2))$ otherwise. 
Note that which design is optimal depends on the sign of the noise asymmetry $\vparam{1}$, which is unknown in advance. 
Furthermore, in either situation, the optimal design cannot extract the actual value of $\vparam{1}$ since the resulting probability distribution depends only on $\vparam{1}+\vparam{2}$ in the case of $(\rho_\pm(1),\Pi(1))$, or $\vparam{1}-\vparam{2}$ in the case of $(\rho_\pm(2),\Pi(2))$.
Therefore, we exclude this singular case.

We next analyze the case for regular designs.
With $\mathrm{Det}(\qfi{\vparam{}}[\rho])\neq 0$, the inverse exists, and
\begin{equation}\label{eq:asymm1}
{\left(\qfi{\vparam{}}[e(1)]^{-1}\right)}_{11}=\frac{1}{4}{\left(\frac{1}{s_1^2}+\frac{1}{s_2^2}\right)}-\vparam{1}^2.
\end{equation}
Eq. \eqref{eq:asymm1} is minimized under the constraint $s_1^2+s_2^2\leq 1$ when $s_1^2=s_2^2=\frac{1}{2}$, taking the value $1-\vparam{1}^2$. 
This SLD CR bound can be saturated by the projective measurement along the directions perpendicular to the optimal input Bloch vector $s_1^2=s_2^2=\frac{1}{2}$ \cite{footnote3}.

Moving onto the $ m = 2 $ case, we look to build a convex structure from rank-1 Fisher information matrices as before. Taking $ \vec{s} $ to be either in the 1 or 2 direction ensures that $ \mathrm{Det}(\qfi{\vparam{}}[\rho])=0 $, i.e., each matrix is rank-1. This suggests a mixed strategy with the two possible input states $ \vec{s}^{(1)} = (s_1, 0, 0)^\mathrm{T} $ and $ \vec{s}^{(2)}= (0, s_2,0)^\mathrm{T}$. The SLD quantum Fisher information of the mixture is the convex sum of the individual Fisher information matrices, 
\begin{equation} \label{eq:asymm.convex.qfi}
\qfi{\vparam{}}[e(2)] = \nu_1 \frac{{s_1}^2\vec{v}_1 \vec{v}_1^\mathrm{T}}{1-{s_1}^2(\vparam{1} + \vparam{2})^2} + \nu_2 \frac{ {s_2}^2\vec{v}_2\vec{v}_2^\mathrm{T}}{1-{s_2}^2(\vparam{1} - \vparam{2})^2},
\end{equation}
where $ \nu_1 + \nu_2 = 1 $. 
Then,
	\begin{align}
	&{\left(\qfi{\vparam{}}[e(2)]^{-1}\right)}_{11}\\
	=& \frac{1}{4s_1^2\nu_1}{\left[1-s_1^2(\vparam{1}+\vparam{2})^2\right]}+\frac{1}{4s_2^2\nu_2}{\left[1-s_2^2(\vparam{1}-\vparam{2})^2\right]} \nonumber,
	\end{align}
which is minimized when {$s^2_1=s^2_2=1$, i.e., $\vec s_1$ and $\vec s_2$ correspond to pure states. We then have, for these pure-state choices of $\vec s^{(1)}$ and $\vec s^{(2)}$,
\begin{align}
\left(\qfi{\vparam{}}[e(2)]^{-1}\right)_{11} &= \frac{f_1^2}{\nu_1} + \frac{f_2^2}{\nu_2}. 
\label{eq:asymm.bin.qfi}
\end{align}

As the inverse of the SLD quantum Fisher information matrix provides a lower bound for the MSE, which is attainable in our setting, we can already compare the lower bounds set by the $ m = 1$ and $m= 2$ designs. Taking the difference between Eqs.~\eqref{eq:asymm1} (with the optimal setting of $s_1^2=s_2^2=\frac{1}{2}$) and \eqref{eq:asymm.bin.qfi}, and setting $\nu_{1,2}= \frac{1}{2}(1\pm\lambda)$ for $\lambda\in[-1,1]$, we have
	\begin{align}
	\Delta{\qfi{\vparam{}}}^{-1} &= \left(\qfi{\vparam{}}[e(1)]^{-1}\right)_{11} - \left(\qfi{\vparam{}}[e(2)]^{-1} \right)_{11}\nonumber \\
	&=1-\vparam{1}^2-{\left[\frac{f_1^2}{\nu_1}+\frac{f_2^2}{\nu_2}\right]}\nonumber\\
	&=\frac{1}{4\nu_1\nu_2}{\left[\vparam{2}^2 - 2\lambda\vparam{1}\vparam{2}-\lambda^2(1-\vparam{1}^2)\right]}.
	\end{align}
For $ \lambda = 0 $, $\Delta{\qfi{\vparam{}}}^{-1} =\vparam{2}^2/4\nu_1\nu_2\geq 0$, i.e., when $\nu_1=\nu_2=\frac{1}{2}$, $ \Delta{\qfi{\vparam{}}}^{-1}$ is nonnegative regardless of the values of $\vparam{1}$ and $\vparam{2}$. This means that for all values of $\vparam{1}$ and $\vparam{2}$, there exists some $\nu_1$ and $\nu_2$ such that $\Delta{\qfi{\vparam{}}}^{-1}  \geq 0$ (e.g., $\nu_1=\nu_2=\frac{1}{2}$).
	From this, we can conclude that it suffices to rule out the $m=1$ case for the optimal design, as $\left(\qfi{\vparam{}}[e(2)]^{-1}\right)_{11}$ can always be equal to or smaller than $ \left(\qfi{\vparam{}}[e(1)]^{-1}\right)_{11} $ from the $m=1$ strategy. 

\begin{figure}
	\centering
	\includegraphics[trim=0mm 0mm 0mm 5mm, clip, width=0.95\columnwidth]{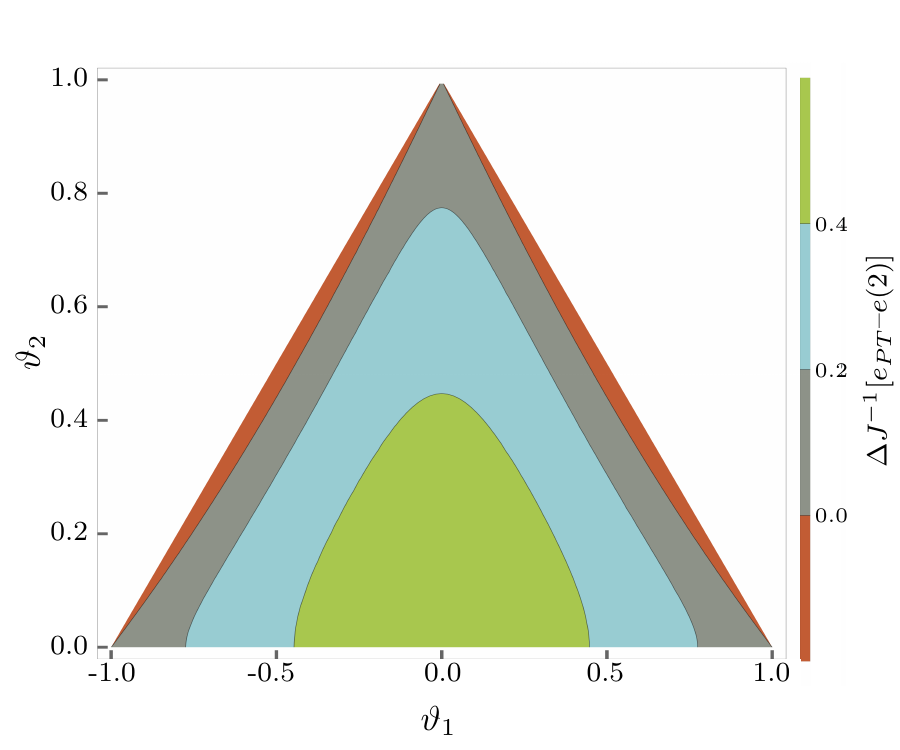}
	\caption{(Color online.) Contour plot of the difference $\Delta J^{-1}[e_{PT}\textrm{--}e(2)]\equiv \left(J_{\vparam{}}[e_{PT}]^{-1}\right)_{11} - 	\left(\qfi{\vparam{}}[e(2)]^{-1}\right)_{11} $ for all possible values of $ \vparam{1}, \vparam{2} $.}
	\label{fig:e3-e2plotlabeled}
\end{figure}

	This does not necessarily mean that $ m =2 $ gives the optimal design, as a larger value of $ m $ could yield an even better design. 
To test its optimality further, we compare $ e(2) $ with a specific $ m = 3 $ case, the Pauli-QPT $ (e_{PT}) $ from Section \ref{sec4-1}. 
	The full calculation of $ \left(J_{\vparam{}}[e_{PT}]^{-1}\right)_{11} $ is rather lengthy, so here we only include the final expression, 
	\begin{equation}
\left(J_{\vparam{}}[e_{PT}]^{-1}\right)_{11}  =3\, \frac{4 f_1^2f_2^2 + f_0^2(f_1^2+f_2^2)}{f_1^2 + f_2^2 + f_0^2},
	\end{equation}}
where $f_0=\sqrt{(1-\vparam{2}) \vparam{2}} $.
We compare this against the $e(2)$ value of $ \left(\qfi{\vparam{}}[e(2)]^{-1}\right)_{11} $ with $ \nu_1 = \nu_2 = 1/2$ by plotting the difference between them; see Fig.~\ref{fig:e3-e2plotlabeled}. 
Pauli-QPT is more optimal only in two narrow slivers of the $\vparam{1}$--$\vparam{2}$ domain along the upper edges of the colored triangle in Fig.~\ref{fig:e3-e2plotlabeled}. These are the regions of extreme asymmetry within the range allowed by the value of $\vparam{2}$.
Since these regions are so small, and it is unlikely that we will have such strong prior information as to expect the $\vparam{1}$ and $\vparam{2}$ values to fall only in those small regions, it is reasonable for us to still consider the $m=2$ design as one that works well, compared with Pauli-QPT. 
This does not, of course, preclude another $m=3$ design from having a larger advantage over the $m=2$ case, or for some larger $ m $ design to be more optimal, as allowed by the Carath{\'e}odory argument. We leave this as an open question for further study.

As was mentioned in the example for the linear scaling channel (Sec.~\ref{sec4-1}), the inequality relating the quantum and the classical Fisher information in the Cram\'er-Rao bound can be saturated by projective measurements along the respective states, which here correspond to the designs $ e_1 $ and $ e_2 $ of Sec.~\ref{sec4-1}. Note that the mixture of $e_1$ and $e_2$ is capable of yielding both channel parameters $\vparam{1}$ and $\vparam{2}$. We thus have a curious case where, even in the presence of a nuisance parameter, a strategy that \emph{fully} characterizes the channel is still the optimal estimation strategy for the noise asymmetry $ \vparam{1}$ alone. 

\subsubsection{Optimal binary design}

To complete the analysis for a continuous design following from the previous section, we must calculate the optimal relative frequencies $ \nu_1, \nu_2 $ for the $m=2$ strategy. Here, we consider the $A$-optimality criterion with the weight matrix $W=\textrm{diag}(1,0)$. With the knowledge that the optimal design is a mixture of $e_1$ and $e_2$, the problem is now equivalent to the binary design problem discussed in Sec.~\ref{sec:twoBin}.
	As before, we write $ \nu_1 = \frac{1}{2}(1 + \lambda) $, $ \nu_2 = \frac{1}{2}(1 - \lambda)$, and we can make use of the formula given in Eq.~(\ref{eq:opt.lambda}). To have a positive weight matrix, we regularize $W$ by setting $ W_\epsilon = \textrm{diag}(1, \epsilon) $ for $\epsilon>0$, and then taking the limit as $\epsilon\rightarrow 0^+$. The function $ \gamma_{\vparam{}}[W](\lambda) $ [see Eq.~\eqref{eq:gamfun.defn}] is given by
\begin{equation}
\gamma_{\vparam{}}(\lambda) = \begin{cases}
f_1^2 & (\textrm{if }\lambda = 1),\\
f_2^2 & (\textrm{if }\lambda = -1), \\
2\left( \frac{f_1^2}{1 +\lambda} + \frac{f_2^2}{1 - \lambda}  \right) & \bigl(\textrm{if }\lambda \in (-1, 1)\bigr).
\end{cases}
\end{equation}
The optimal partition can be found by a derivative test or by using Eq. (\ref{eq:opt.lambda}), where $ \lambda_{\pm} = \pm 1 $. The optimal $ \lambda $-value is then
\begin{equation}\label{eq:optL}
\lambda^{*}=\frac{f_1-f_2}{f_1+f_2}.
\end{equation}
Let $e_*(2)$ be the local optimal design with this choice of frequency $\v{\nu}_*=(\frac{1+\lambda^*}{2},\frac{1-\lambda^*}{2})$.
The minimum value for the Quantum Fisher Information, and also the MSE matrix for a locally unbiased estimator, is then 
\begin{equation}
(J_{\vparam{}}[e_*(2)]^{-1})_{11}={\left(f_1+f_2\right)}^2.
\end{equation}
From the above derivation, we again confirm that the singular design \eqref{eq:loc_sing_opt} is 
the local optimal design, since ${\left(f_1+f_2\right)}^2\ge \min\{f_1^2,f_2^2\}$ always holds.

\begin{figure}
	\centering
	\includegraphics[trim=0mm 5mm 0mm 0mm. clip, width=0.9\columnwidth]{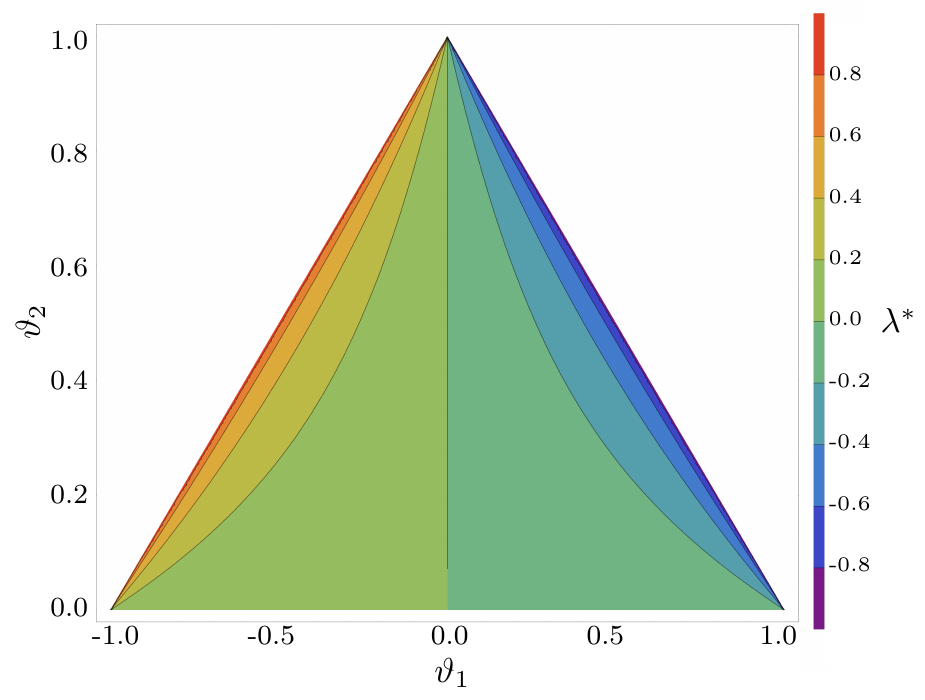}
	\caption{(Color online.) A contour plot of $ \lambda^{*} $ [Eq. (\ref{eq:optL})] as a function of $ \vparam{1} $ and $ \vparam{2} $.}
	\label{fig:optlam}
\end{figure}

The optimal value $\lambda^*$ depends on the unknown parameters $\vparam{1}$ and $\vparam{2}$. Without \emph{a priori} knowledge of $\vparam{1}$ and $\vparam{2}$, one cannot implement the optimal design.
However, observe in Fig.~\ref{fig:optlam}, a contour plot of $\lambda^*$ in the $\vparam{1}$--$\vparam{2}$ domain, that the magnitude $|\lambda^*|$ is relatively small and flat over a large central region and rises sharply only in the high asymmetry regions. 
This hints at the possibility of an adaptive approach for better performance: We can start with $\lambda^*=0$, or, equivalently, with equal weights on $e_1$ and $e_2$, and then adapt the relative weights towards the optimal $\lambda^*$ as we gather information about the actual values of $\vparam{1}$ and $\vparam{2}$. We expect this adaptation to be particularly important for the high asymmetry regions. 

To understand how much benefit we can gain, we examine this adaptive strategy in the next section. We will, however, use a discrete, rather than continuous, design, so that one can look at the performance with finite data, instead of the mathematical asymptotic limit.
To make this transition to a discrete design, we consider a strategy 
$ e(N,\lambda) $
where a fixed number $N_1\equiv \frac{1}{2}(1+\lambda) N$ of uses of the channel is for the $e_1$ design and $N_2\equiv \frac{1}{2}(1-\lambda)N$ is for the $e_2$ one, for a total $N=N_1+N_2$ uses. $\lambda$ is now regarded as a parameter that characterizes the \emph{fixed fraction}, rather than the probability, of the $N$ uses of the channel that employs design $e_1$ or $e_2$, and $ \lambda^{*} $ is the optimal fixed fraction. We will assume here that $N_1,N_2\geq 1$.

As before, we are interested only in estimating $\vparam{1}$. Let $n_i$ denote the number of counts entering the detector for $\Pi_{i,+}$, out of $N_i$ counts that used design $e_i$, for $i=1,2$. We construct the estimator $\hvparam{1}$ for $\vparam{1}$ as
\begin{equation} 
\hvparam{1}\equiv   \frac{n_1}{N_1} - \frac{n_2}{N_2},
\end{equation}
For given $N_1$ and $N_2=N-N_1$, the MSE is simply, from its definition [see Eq.~\eqref{eq:mse.matrix}], 
\begin{equation}
V_{\vparam{}}[\hvparam{1}|e(N,\lambda)] = \frac{f_1^2}{N_1}+\frac{f_2^2}{N_2}.
\end{equation}
Compared to the SLD quantum Fisher information of Eq.~\eqref{eq:asymm.bin.qfi}, there is an additional factor of $\frac{1}{N}$ to account for the $N$ uses of the channel.

\subsubsection{Adaptive discrete design}
As in the continuous design case, one can expect the optimal discrete design $e(N,\lambda^*)$ to depend [see Eq.~\eqref{eq:optL}] on the values of $f_1$ and $f_2$. These in turn depend on the unknown values of $\vparam{1}$ and $\vparam{2}$. As argued above, one expects an adaptive strategy to to be helpful in such a situation.
We implement one such strategy by dividing the total available uses of the channel $N$ into $K$ adaptive steps. In each step we decide on the relative proportion of $e_1$ and $e_2$ using estimates of $f_1$ and $f_2$ from the data gathered so far, up to the last completed step.

Specifically, we let $N=M_1+M_2+\ldots+M_K$, where $M_k$, for $k=1,\ldots, K$, is the number of uses of the channel in the $k$th adaptive step. Let $\lambda_k$ be the $\lambda$ parameter for the $k$th step, i.e., we devote $N_{k,1}\equiv \frac{1}{2}(1+\lambda_k)M_k$  uses to the $e_1$ design and $N_{k,2}\equiv\frac{1}{2}(1-\lambda_k)M_k$ to $e_2$. We denote the total number $\sum_{\ell=1}^kN_{\ell,i}$ of uses for $e_i$ so far, up to and including the $k$th step, by $N_{1:k,i}$, for $i=1,2$. We further define $n_{k,i}$  to be the number of detector clicks for $\Pi_{i,+}$ in the $k$th step. Analogously, $n_{1:k,i}$ denotes the number of detector clicks for $\Pi_{i,+}$ so far, up to the $k$th step. After $k$ adaptive steps, we estimate $f_i$, for $i=1,2$, by
\begin{equation}
\widehat f_{k,i}= {\left[\frac{n_{1:k,i}}{N_{1:k,i}}{\left(1-\frac{n_{1:k,i}}{N_{1:k,i}}\right)}\right]}^{1/2}.
\end{equation}
We use these estimates of $f_1$ and $f_2$ to determine the optimal $\lambda_{k+1}$ for the next adaptive step. Eq.~\eqref{eq:optL} suggests that the optimal $\lambda$ for the \emph{total} number of uses of the channel for all $k+1$ steps, is 
\begin{equation}
\lambda_{1:k+1}= \frac{\widehat f_{k,1}-\widehat f_{k,2}}{\widehat f_{k,1}+\widehat f_{k,2}}.
\end{equation}
From this, we see that the optimal choice for $N_{k+1,1}$, say, in the next adaptive step, is given by
\begin{equation}
N_{k+1,1}={\left[\tfrac{1}{2}(1+\lambda_{1:k+1})M_{1:k+1}-N_{1:k,1}\right]}_+,
\end{equation}
where the notation $y=[x]_+$ is shorthand for $y=x$ when $x\geq 0$, and $y=0$ when $x<0$. Straightforward algebra gives
\begin{equation}
\lambda_{k+1}=2{\left[\frac{\widehat f_{k,1}+\widehat f_{k,1}\frac{N_{1:k,1}}{M_{k+1}}-\widehat f_{k,2}\frac{N_{1:k,2}}{M_{k+1}}}{\widehat f_{k,1}+\widehat f_{k,2}}\right]}_+-1.
\end{equation}
To start the adaptive sequence, we need to decide on the initial estimates of $f_1$ and $f_2$. With no prior information, reasonable initial guesses for $\vparam{1}$ and $\vparam{2}$ are $0$ and $\frac{1}{2}$, respectively, the midpoints of the allowed ranges of $\vparam{1}$ and $\vparam{2}$. This corresponds to initial estimates 
$\widehat f_{0,1}=\widehat f_{0,2}=\tfrac{\sqrt 3}{4}$, and a vanishing starting value of $\lambda_1$, i.e., $M_1$ is divided equally between $e_1$ and $e_2$.

We compare the MSE of an adaptive scheme with that of a static, i.e., non-adaptive, strategy where the $N$ uses of the channel are shared equally between $e_1$ and $e_2$. Let $\lambda_\mathrm{eff}\equiv \frac{2N_{1:K,1}}{N}-1$ be the effective $\lambda$ parameter for the adaptive scheme. The relative performance of the two schemes, for the same channel (i.e., fixed values of $f_1$ and $f_2$), can be written as
\begin{equation}\label{eq:Vratio}
\frac{V_\mathrm{static}}{V_\mathrm{adapt}}=(1-\lambda_\mathrm{eff}^2){\left[-\lambda_\mathrm{eff}\frac{f_1^2-f_2^2}{f_1^2+f_2^2}+1\right]}^{-1}.
\end{equation}

\begin{figure}
\centering
\includegraphics[trim=1mm 1mm 3mm 0mm, clip, width=\columnwidth]{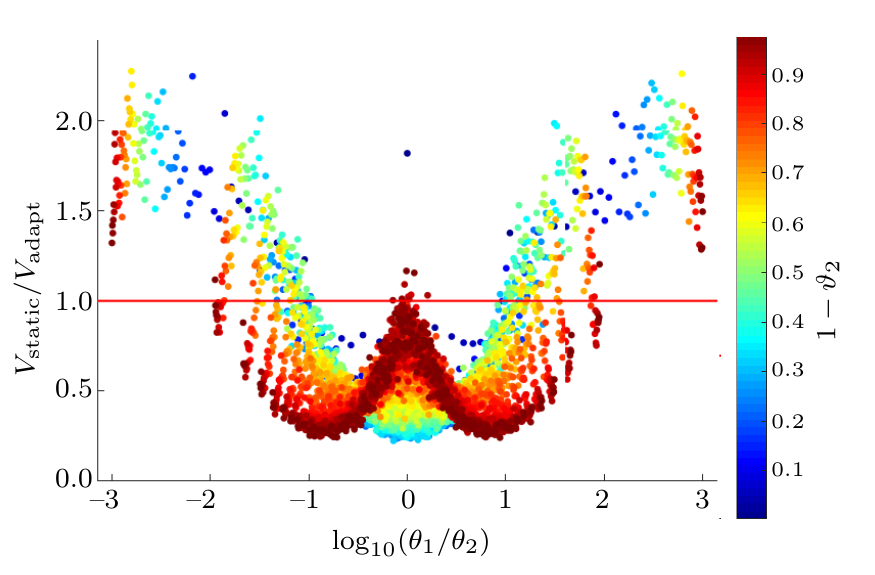}
\caption{\label{fig:magnitude.split} (Color online.) A plot of the ratio $V_\mathrm{static}/V_\mathrm{adapt}$ against $\log_{10}(\param{1}/\param{2})$ for numerical simulation of the adaptive procedure for the two-parameter family of Pauli channels, over a uniform grid of $\param{1}$ and $\param{2}$ values, with a step size of $0.01$. Here, $N=200$, and $K=10$. The colors show the dependence of the MSE ratio on the noise strength $1-\vparam{2}$; a plot of only the data points for the low-noise regime of $1- \vparam{2} \leq 0.5$ is given in Fig.~\ref{fig:runway.comparison}(a) below. That the adaptive strategy shows a clear benefit in the regime of high asymmetry is evidenced by the points that lie above the $V_\mathrm{static}/V_\mathrm{adapt}=1$ horizontal line, which occur only when $\param{1}/\param{2}$ is far from 1.}
\end{figure}
	
\begin{figure*}
\centering
\includegraphics[trim=0mm 0mm 0mm 10mm, clip, width=0.9\textwidth]{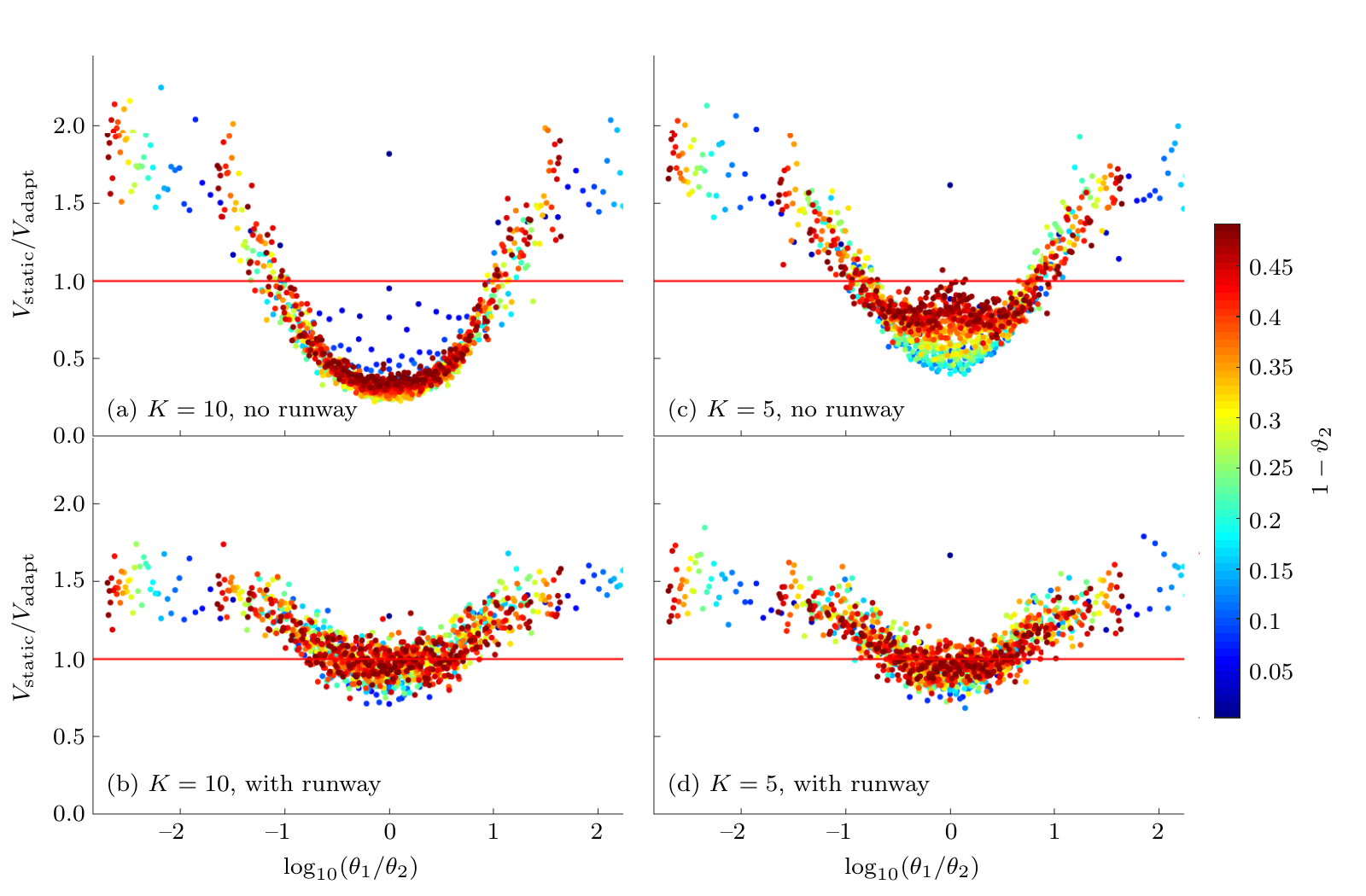}
\caption{\label{fig:runway.comparison}(Color online.) The effect of a runway on the MSE ratios. We have a total of $N=200$ uses of the channel, and only the $1-\vparam{2}\leq 0.5$ data are shown here. (a) $K=10$, no runway; (b) $K=10$, a runway of length 100; (c) $K=5$, no runway; (d) $K=5$, a runway of length 100. The runway clearly reduces the loss in accuracy due to the adaptation when $\param{1}/\param{2}$ is close to 1.}
\end{figure*}

To examine the performance of the adaptive procedure, we carried out numerical experiments for an estimation scheme with $N=200$, and $K=10$ for the two-parameter family of Pauli channels. The simulations were run over a uniform grid of all possible $ \param{1}$ and $\param{2}$ values [recall the relationship between $\vparam{i}$s and $\param{i}$s, as given in Eq.~\eqref{eq:noise.params}], with a step size of $0.01$. 
The results are given in Fig.~\ref{fig:magnitude.split}, which plots the ratio $V_\mathrm{static}/V_\mathrm{adapt}$ against $\log_{10}(\param{1}/\param{2})$, a quantity we found useful in organizing the data. 
The MSE ratios in Fig.~\ref{fig:magnitude.split} show a dependence on the value of noise strength $1-\vparam{2}$, as indicated by the colors in the plot. Clearly visible in the plot is the strong benefit of using the adaptive strategy in a high asymmetry situation: The ratio $V_\mathrm{static}/V_\mathrm{adapt}$ is large when when $\param{1}/\param{2}$ is very different from 1, i.e., when $|\log_{10}(\param{1}/\param{2})|$ is large.

Away from the high asymmetry region, however, the adaptive scheme in fact does more poorly than the static scheme, except for the rare case that $ \vparam{1}=\vparam{2} = 0$. It is not difficult to guess why this might be the case. In this region, the optimal value of $\lambda^*$ should stay close to $0$, as seen from Fig.~\ref{fig:optlam}. However, in the early phase of the experiment when one does not have a lot of data, statistical fluctuations can easily cause the adaptive scheme to opt for $\lambda$ values that are away from the optimal $0$ value. The adaptive scheme hence may meander around initially before we gather enough data to have good guidance in the adaptation, while the static scheme is already using a near-optimal $ \lambda $ value.

To mitigate the effects of this initial meandering, we can modify our adaptive scheme to include an initial ``runway", where an initial number of measurements are made without any adaptations, using a fixed equal weight between the $e_1$ and $e_2$ designs. The adaptation kicks in only when we have gathered enough data. The effect of a runway is shown in the numerical simulation data given in Fig.~\ref{fig:runway.comparison}. For quantum information processing applications, one is usually interested only in the low-noise regime, say, where $1- \vparam{2} \leq 0.5$, so we focus only on the data points in this regime. As before, we have a total of $N=200$ uses of the channel, and only the $1-\vparam{2}\leq 0.5$ data are shown in the plots. The MSE ratios are shown for four situations: (a) $K=10$, no runway (this is the same data as that of Fig.~\ref{fig:magnitude.split}, but restricted to the regime of $1- \vparam{2} \leq 0.5$); (b) $K=10$, a runway with $N/2=100$ uses of the channel, followed by the remaining $N/2=100$ uses for adapatation (in the adaptive scheme); (c) as in (a) but now with $K=5$; (d) as in (b), but now with $K=5$.

The runway clearly helps reduce the loss in accuracy due to the adaptation for $\param{1}/\param{2}\approx 1$, compared to the static scheme. However, it also reduces the edge of the adaptation over the static case in the high asymmetry regime, as can be expected given the fewer channel uses available for adaptation.
Another feature visible in Fig.~\ref{fig:runway.comparison} is that the region (of $\param{1}/\param{2}$ values) where adaptation helps shrinks slightly when $K$ is reduced from 10 to 5, as does the range of the MSE ratio values. A full exploration of the adaptation strategy should also investigate the optimal values for $K$ and the length of the runway. The optimal choice of runway length will no doubt depend on the available prior information about the noise asymmetry in the channel. Note that, in our numerical simulations, we did not find a combination of runway length and $K$ values that shrank the region where $V_\mathrm{static}/V_\mathrm{adapt}<1$ to zero.
\textsl{}

Let us make a final remark about the noise asymmetry example. The noise asymmetry of the \emph{full} Pauli channel of Example B can also be described via a suitable parameterization. In that case, two asymmetry parameters can capture the relative strengths of the three original parameters, requiring a nuisance parameter to fully characterize the Pauli channel. We can define new parameters $ \vparam{1} = \param{1} - \param{2} $ and $ \vparam{2} = \param{3} - \param{2} $, 
$\vparam{3}=1 - (\param{1} + \param{2} + \param{3}) $. The parameters of interest here are the asymmetry parameters $\vparam{1}$ and $\vparam{2}$; the nuisance parameter is $ \vparam{3}$. Preliminary numerical simulation of the full Pauli channel situation indicate similar observations that the full characterization of three-parameter Pauli channel is near optimal even in the presence of the nuisance parameter, i.e., the design $ J_{\vparam{}}[e(3)] = \sum_{i = 1,2, 3} \nu_i J_{\vparam{}}[e_i] $ is near optimal.

\section{Summary and outlook}

In summary, we formulated the problem of quantum process tomography within the framework of optimal design of experiments (DoE). This allows us to adapt the many techniques developed in classical statistics to quantum tomography problems, as demonstrated in the examples discussed in this paper. Here, we worked out simple examples to get analytical results for clearer illustration; more generally, the question of finding an optimal design can be solved efficiently as a convex optimization problem. 

One of the well-known issues in standard optimal DoE problems is that one often finds a local optimal design. 
This local optimal design normally depends on the values of unknown parameters, including the very ones we are trying to estimate. Such a design hence cannot be realized exactly in practice. This point was demonstrated by several examples of qubit noise channels. 

The standard remedy against this local optimal design problem is to utilize an appropriate adaptive scheme. 
This is a well-established strategy known in the community. 
In the example of detecting noise asymmetry for the Pauli channel (Sec.~\ref{sec5-3}), we applied a particular adaptive scheme by splitting $N$ uses of the channel into $K$ steps of adaptation together with the use of a runway stage to acquire some information about the unknown parameters. From our numerical simulations, we observed a gain from this particular adaptive scheme in a high asymmetry regime. In the low asymmetry regime, however, the adaptive scheme did worse than the static one. This is partly due to the fact that the static design (of $\lambda^*=0$) is near optimal for wide ranges of parameters (see Fig.~\ref{fig:optlam}). This conclusion is, of course, highly dependent on the specifics of our particular example, but this raises an important question about the effectiveness of adaptation for implementing local optimal designs, worthy of further investigation.

The same noise asymmetry example of Sec.~\ref{sec5-3} also raises another important point about nuisance parameters and singular designs. It is often stated in DoE literature that the generalized inverse sets the bound for the mean square error of estimates. While this is a correct mathematical statement, the actual realization of such a singular design has to be carefully examined before one can claim its optimality. This is particularly important when dealing with problems in the presence of nuisance parameters. In our example, the singular design simply cannot be used to estimate the noise asymmetry, the quantity of interest. Instead, the static design that actually estimates the noise asymmetry turns out to be one that can estimate all the parameters of the problem, i.e., both the quantity of interest and the nuisance parameter. Of course, since we have only investigated strategies up to $m=2$, we cannot claim the nonexistence of a strategy with larger $m$ capable of efficiently estimating only the parameter of interest. Yet, a small-$m$ strategy is one of most interest in practical implementations.

In this work, we have but scratched the surface of the many possibilities of adopting ideas from the rich subject of classical theory of DoE. We expect further exploration in this direction to yield useful and interesting results for the quantum problem of process tomography.

\acknowledgments
This work is supported in part by the Ministry of Education, Singapore (through grant number MOE2016-T2-1-130). The Centre for Quantum Technologies is a Research Centre of Excellence funded by the Ministry of Education and the National Research Foundation of Singapore. JS is partly supported by JSPS KAKENHI Grant Number JP17K05571. HKN is also supported by Yale-NUS College (through a start-up grant).

\appendix*
\section{Supplemental material}
	
\subsection{Positive-definite matrix}
	We denote the set of all real positive-definite and positive-semidefinite matrices of size $n$ by  $\pd$ and $\nnd$, respectively. 
	It is known that the following conditions are equivalent (see for example, Ref.~\cite{bhatia}),
	\begin{align} \label{eq:pdcond}
 A \in \pd&\DEF \forall v\in\bbr^n,\, v\neq\v{0}\Rightarrow v^\mathrm{T}Av>0\\ \nonumber
&\Lra \forall B\in \nnd,\, B\neq\v{0}\Rightarrow\Tr{AB}>0,\\ \label{eq:nndcond}
 A \in \nnd&\DEF \forall v\in\bbr^n,\, v^\mathrm{T}Av\ge0\\ \nonumber
&\Lra \forall B\in \nnd,\, \Tr{AB}\ge0.
\end{align}

\subsection{Optimality function $\Psi$}\label{sec-app_supp1}
	Mathematically, any function from a positive semidefinite matrix space to a real positive number, $\Psi:$ NND$(n)\to\bbr$ 
	can be used as the optimality function. 
	From statistics and information theoretical view, we normally impose the following properties on $\Psi$: 
	\begin{enumerate} 
		\item Isotonicity (operator monotone function): \\For $J_1,J_2\in\nnd$, if $J_1\ge J_2$, then $\Psi(J_1)\le\Psi(J_2)$. 
		\item Homogeneity: There exists a function $\psi$ such that $\Psi(a J)=\psi(a) \Psi(J)$ holds for any constant $a>0$ and all $J\in\nnd$.
		\item Convexity: $\Psi(pJ_1+(1-p)J_2)\le p \Psi(J_1)+(1-p) \Psi(J_2)$, for $\forall p\in[0,1]$ and $\forall J_1,J_2\in\nnd$.
	\end{enumerate}
	Note that the popular optimality criteria discussed in the main text are expressed as follows. 
	$A$-optimality: $\Psi_A(J)=\Tr{J^{-1}}$, $D$-optimality: $\Psi_D(J)=\Det{(J^{-1})}$, 
	$E$-optimality: $\Psi_E=\max_{c\in\bbr^n}c^\mathrm{T}J^{-1}c/|c|^2$, $c$-optimality: $\Psi_c=c^\mathrm{T}J^{-1}c$. 
	It is easy to check that $A$- and $E$-optimality satisfy the above three conditions. 
	$D$-optimality violates the third condition, however, and the standard remedy is to instead optimize $\log  \Det{J_{\param{}}[e]^{-1}}=- \log  \Det{J_{\param{}}[e]}$, which is a convex function. 	
	
As noted in the main text, the $\gamma$-optimality criterion contains the $A$-optimal ($\gamma=1$), $D$-optimal ($\gamma\to0$), and $E$-optimal ($\gamma\to\infty$) criteria as special cases up to appropriate constant factors. In this sense, the $\gamma$-optimality 
is a sort of generalization of the standard optimality criteria. 

There is a non-trivial inequality relation among the \mbox{$A$-,} $D$-, and $E$-optimality functions \cite{fh97,fl14}. 
A well-known Liapunov's inequality in probability theory proves
\begin{equation} \label{eq:Liapunov.ineq}
(\Det{J^{-1}})^{1/n}\le\frac1n\Tr{J^{-1}}\le\lambda_{\max}(J^{-1}),
\end{equation}
where $n$ is the dimension of the parameter set $\Param$. 
These inequality relationships, however, do not provide a general hierarchy among these criteria. 
	
	Two optimality functions (or, more generally, many optimality functions) can be combined to define a {\it compound optimality function}  $\Psi_p=p \Psi_1+(1-p) \Psi_2$ with $p \in [0, 1]$. 
	$\Psi_{p}$  represents a trade-off relation between two different optimal designs defined by $\Psi_1$ and $\Psi_2$. 
	
	Careful consideration is needed to compare different optimality functions. 
	The value of $\Psi[e_*]$ is a relative quantity, and we cannot conclude that an $e_*$ for a particular $\Psi$ is also a good design according to another optimality function $\Psi'$ simply by looking at the value of $\Psi'[e_*]$. 
	To compare the performance of different optimality criteria, one can consider a function $ \eta $ dependent on the optimal value $\Psi[e_*]$, ${\eta_\Psi[e]=\Psi[e_*]/\Psi[e]}$. 
	The normalized function satisfies $0\le\eta_\Psi[e]\le1$, and it can describe the efficiency of the design $e$ for the function $\Psi$. 
	Applications of the above extended optimal designs were discussed in various statistical problems, see Refs.~\cite{pukelsheim,fh97,fl14,pp13}. 
	
\subsection{L\"owner optimal design} \label{sec-app_lowner}
One way to check if a L\"owner optimal design exists is to minimize a weighted $A$-optimal function $ \Psi_{A}(J) = \Tr{ W J^{-1} } $.
	If the optimal design $ e_* $ is $ W $-independent, then a L\"owner optimal design exists. 
	Otherwise, its existence is disproved. The reason behind this logic is the expression for the positive semidefinite matrix 
	given by \eqref{eq:nndcond}.
	Alternatively, one can work out the $c$-optimal design problem and then to check if the optimal design is independent of $c$.

	It is pointed out in the main text that the L\"owner optimality is the strongest criterion. 
	That is, if there exists a L\"owner optimal design $e_*$, then $e_*$ is also optimal 
	for all the other optimality criteria. To show this, let us assume that the optimality function $\Psi$ satisfies
	the three conditions discussed in Sec.~\ref{sec-app_supp1}. In particular, when $\Psi$ is isotonic, 
	then, for $J_{\param{}}[e_*]\ge J_{\param{}}[e]$, we have,
	\begin{equation}
	\forall e\in\cE,\ \Psi(J_{\param{}}[e_*])\le \Psi(J_{\param{}}[e]). 
	\end{equation}
	This shows that the L\"owner optimal design $e_*$ is indeed an optimal design for the optimality 
	criterion defined by $\Psi$.

\subsection{Convex optimal structures}\label{sec-app_supp2}	
	A common approach to finding an optimal design $ e_* $ is to introduce a convex structure to the design problem. 
	This allows for a systematic search over the entire parameter space for an optimal design. 
	Two separate ingredients are needed to create this convexity. 
	The first is a convex structure on the design set $\cE$: A convex sum of two designs $e_1,e_2\in\cE$ is defined as ${e_p=p e_1+(1-p)e_2}$ for $p\in[0,1]$. 
	It is easy to show that this convex structure preserves the locally unbiasedness of an estimator.
	
	Having this convex structure for the design space $\cE$ and convexity of the function $\Psi$, 
	we can formulate our problem as a convex optimization problem. 
	An important consequence of this formulation is that such a convex problem has optimal designs $e_*$ at the extremal points of the convex set $\cE$. 
	The necessary and sufficient condition for such an optimal design can also be derived; see, for example, Refs.~\cite{fedorov,pukelsheim,fh97,fl14,pp13}. 

	Remember that the continuous design problem aims to find an optimal design $e_*(m)=(\v{\nu}_*,\v{e}_*)$ defined by 
	\begin{equation}\label{eq:opt.design2}
	e_*(m)=\arg\hspace{-5mm}\min_{e(m)\in\cP(m)\times\cE^m} \Psi\bigg(\sum_{i=1}^m \nu_iJ_{\param{}}[e_i]\bigg). 
	\end{equation}
	A convex structure can also be constructed from two continuous designs $e(m)=(\v{\nu},\v{e})$ and $e'(m)=(\v{\nu}',\v{e}')$,  
	\begin{equation*}
	p e(m)+(1-p) e'(m)=\big(p\v{\nu}+(1-p)\v{\nu}',\, p\v{e}+(1-p)\v{e}' \big),
	\end{equation*}
	where $p\v{e}+(1-p)\v{e}' =\big(pe_i+(1-p)e'_i \big)$ is a well-defined convex sum of two designs. 
	
	Special consideration is needed to define a convex sum of two designs $e(m)$ and $e'(m')$ for $m\neq m'$.	This is done by introducing an integration measure $\mu$  for the design space $\cE$, allowing one to consider an experimental design of the form $e_\mu=\int \mu(de) e$. 
	This formalism is more general, since a discrete measure contains the above-mentioned continuous design problem as a special case. 
	In the literature on mathematical theory of optimal DoE, optimization of the measure $\mu$ is studied; see, for example, Refs.~\cite{fedorov,pukelsheim,fh97,fl14,pp13}. 
	
	Another important problem is characterizing the structure of the Fisher information matrices for all possible designs, 
	i.e., finding a set of matrices defined by
	\begin{equation*}
	\cJ(m)= \{J=\sum_{i=1}^m \nu_iJ_{\param{}}[e_i]\,|\, \v{\nu}\in\cP(m),e_i\in\cE \}, 
	\end{equation*}
	or more generally, their unions
	\begin{equation*}
	\cJ=\bigcup_{m\in\bbn}\cJ(m). 
	\end{equation*}
	Then, the optimization problem can be rephrased as a minimization over the convex set,
	\begin{align}\nonumber
	\Psi_*&=\min_{J\in\cJ} \Psi(J),\\ 
	J_{\param{}}[e_*]&=\arg\min_{J\in\cJ} \Psi(J). 
	\end{align}

\subsection{Singular design}\label{sec-app_sing}
	
A design $e\in\cE$ is called a {\it singular design} if the Fisher information matrix $J_{\param{}}[e]$ is not full rank. 
	When $J_{\param{}}[e]$ is invertible, $ e $ is a {\it regular design}. 
	A typical example of an optimal singular design is the $c$-optimal design, which corresponds to finding an optimal design in a particular direction of the information matrix. 
	Such an optimal design should be in the direction specified by the c-vector, giving a singular Fisher information matrix. 
	
	The inverse of the Fisher information matrix does not exist for singular designs, so an appropriate remedy is needed for a properly defined optimal design. 
	The most common approach is to use the generalized inverse of the Fisher information matrix. 
	In particular, the Moore-Penrose inverse matrix is uniquely defined for singular matrices, and is used in literature; see for example the standard textbook \cite{lc98} and also Ref.~\cite{sm01} on this issue. 
	A common alternative is the regularization method: Given a singular matrix $J$, we calculate $(J+\epsilon I)^{-1}$ with $\epsilon>0$, and then take the limit $\epsilon\to0$. 
	These two methods, however, are not always practical, in which case more alternatives are needed. 
	
Yet another approach is to find an $A$-optimal design for the following positive-definite weight matrix, 
\begin{equation}
W_\epsilon=\left(\begin{array}{cc}W_{I} & 0 \\ 0& \epsilon I_N\end{array}\right), 
\end{equation}
where $I_N$ is the identity matrix for the $(n-k)\times(n-k)$ sub-block. 
Under mild regularity conditions, the optimal design ${e_* = \arg\min\Tr{W_\epsilon J_{\param{}}[e]^{-1}}}$ exists and is regular. 

Or, one can optimize a {\it regularized} optimality function by creating a continuous design problem. 
	Let $e_0$ be a design such that $J_{\param{}}[e_0]$ is regular. 
	Then, we can regularize any optimality function by using the Fisher information matrix $(1-\epsilon)J_{\param{}}[e]+\epsilon J_{\param{}}[e_0]$ 
	for any $e\in\cE$ and $\epsilon\in(0,1)$. 
	
	Lastly, one can use a compound optimal criterion. 
	Let $\Psi$ be the optimality function under consideration whose optimal design can be singular. 
	We can consider another optimality function $\Psi'$ such that $e_*=\arg\min\Psi'(e)$ is always regular. 
	The combined optimality function $\Psi_\epsilon=(1-\epsilon) \Psi+\epsilon \Psi'$ (discussed in Appendix Sec.~\ref{sec-app_supp1}) 
	can then be used to define a regular optimal design.




\begin{thebibliography}{20}
\bibitem{pauli}
		The standard representation of the Pauli matrices is 
		{\scriptsize $\sigma_1=\left(\begin{array}{cc}0 & 1 \\1 & 0\end{array}\right)$, 
			$\sigma_2=\left(\begin{array}{cc}0 & -\I \\ \I & 0\end{array}\right)$, 
			$\sigma_3=\left(\begin{array}{cc}1 & 0 \\0 & -1\end{array}\right)$.}
\bibitem{footnote2}
		To see this, note that $|\vparam{1}|=\param{>}-\param{<}$, where $\param{>(<)}$ is the larger(smaller) of $\param{1}$ and $\param{2}$. Then, $\vparam{2}=\param{>}+\param{<}=|\vparam{1}|+2\param{<}\geq |\vparam{1}|$ since $\param{<}\geq 0$.
\bibitem{footnote3}
		If the input state is $\v{s}=\frac{1}{\sqrt{2}}( 1,1,0)^\mathrm{T}$ or $\frac{1}{\sqrt{2}}( -1,-1,0)^\mathrm{T}$, 
		the corresponding projective measurement is $\Pi=\{ \frac{1}{2}[I_d\pm \frac{1}{\sqrt{2}}(\sigma_1-\sigma_2)]\}$. 
		For the input state $\v{s}=\frac{1}{\sqrt{2}}( 1,-1,0)^\mathrm{T}$ or $\frac{1}{\sqrt{2}}( -1,1,0)^\mathrm{T}$, 
		the corresponding projective measurement is $\Pi=\{ \frac{1}{2}[I_d\pm \frac{1}{\sqrt{2}}(\sigma_1+\sigma_2)]\}$.
\end{thebibliography}
\end{document}